\documentclass[article,pre]{revtex4}
\usepackage{graphicx}
\usepackage{epsfig}
\usepackage{amsmath,amssymb}
\usepackage{graphics}
\usepackage{latexsym}
\usepackage{microtype}
\begin{document}
\title{Monte Carlo Tests of Nucleation Concepts in the Lattice Gas Model}

\author{Fabian Schmitz, Peter Virnau and Kurt Binder}

\affiliation{Institute of Physics, Johannes Gutenberg-Universit\"at Mainz, Germany}

\begin{abstract}
The conventional theory of homogeneous and heterogeneous nucleation in a supersaturated vapor is tested by Monte Carlo simulations of the lattice gas (Ising) model with nearest-neighbor attractive interactions on the simple cubic lattice. The theory considers the nucleation process as a slow (quasi-static) cluster (droplet) growth over a free energy barrier $\Delta F^*$, constructed in terms of a balance of surface and bulk term of a ``critical droplet'' of radius $R^*$, implying that the rates of droplet growth and shrinking essentially balance each other for droplet radius $R=R^*$. For heterogeneous nucleation at surfaces, the barrier is reduced by a factor depending on the contact angle. Using the definition of ``physical'' clusters based on the Fortuin-Kasteleyn mapping, the time-dependence of the cluster size distribution is studied for ``quenching experiments'' in the kinetic Ising model, and the cluster size $\ell ^*$ where the cluster growth rate changes sign is estimated. These studies of nucleation kinetics are compared to studies where the relation between cluster size and supersaturation is estimated from equilibrium simulations of phase coexistence between droplet and vapor in the canonical ensemble. The chemical potential is estimated from a lattice version of the Widom particle insertion method. For large droplets it is shown that the ``physical clusters'' have a volume consistent with the estimates from the lever rule. ``Geometrical clusters'' (defined such that each site belonging to the cluster is occupied and has at least one occupied neighbor site) yield valid results only for temperatures less than 60\% of the critical temperature, where the cluster shape is non-spherical. We show how the chemical potential can be used to numerically estimate $\Delta F^*$ also for non-spherical cluster shapes.
\end{abstract}
\maketitle

\section{Introduction}
 \label{sec1}
Since the theory of nucleation phenomena was introduced a long time ago \cite{1,2,3}, the question under which conditions the ``conventional theory'' of nucleation is accurate has been debated (see e.g. \cite{4,5,6,7,8,9,10,11,12,13,14,15,16,17,18,19,20,21,22,23,24}) and this debate continues until today. For the simplest case of homogeneous nucleation (by statistical fluctuations in the bulk) of a one-component liquid droplet from the vapor, the basic statement of the theory is that under typical conditions nucleation processes are rare events, where a free energy barrier $\Delta F^*$ very much larger than the thermal energy $k_BT$ is overcome, and hence the nucleation rate is given by an Arrhenius law,
	\begin{equation}\label{eq1}
	j = \omega \exp (-\Delta F ^*/k_BT)\; .
	\end{equation}
Here $j$ is the number of nuclei, i.e. droplets that have much larger radii $R$ than the critical radius $R^*$ associated with the free energy barrier $\Delta F^*$ of the saddle point in configuration space, that are formed per unit volume and unit time; $\omega$ is a kinetic prefactor. Now $\Delta F^*$ is estimated from the standard assumption that the formation free energy of a droplet of radius $R$ can be written as a sum of a volume term $(\propto 4 \pi R^3/3)$, and a surface term $(\propto 4 \pi R^2)$, i.e.
	\begin{equation}\label{eq2}
	\Delta F(R) = - \frac{4\pi R^3}{3} \Delta \mu (\rho_\ell - \rho_v) + 4 \pi R^2 \gamma _{v\ell}\; .
	\end{equation}
Since the liquid droplet can freely exchange particles with the surrounding vapor, it is natural to describe its thermodynamic potential choosing the chemical potential $\mu$ and temperature $T$ as variables, and expand the difference in thermodynamic potentials of liquid and vapor at the coexistence curve, $\Delta \mu = \mu - \mu_\text{coex}$, $\rho_v$ and $\rho_\ell$ denoting the densities of the coexisting vapor ($v$) and liquid ($\ell$) phases. According to the capillarity approximation, the curvature dependence of the interfacial tension $\gamma_{v\ell}$ is neglected, $\gamma_{v\ell}$ is taken for a macroscopic and flat vapor-liquid interface. Then the critical radius $R^*$ follows from
	\begin{equation}\label{eq3}
	\left.\frac{\partial \Delta F(R)}{\partial R}\right|_{R^*}=0, \quad R^* = \frac{2 \gamma_{v \ell}}{\Delta \mu(\rho_\ell - \rho_v)}\;,
	\end{equation}
and the associated free energy barrier is
	\begin{equation}\label{eq4a}
	\Delta F^*_\text{hom} = \frac{4\pi}{3} (R^*)^2 \gamma_{v \ell}\;.
	\end{equation}
However, since typically $\Delta F^*_\text{hom}$ is less than 100 $k_BT$, the critical droplet is a nanoscale object, and thus the treatment Eqs.~\eqref{eq1}-\eqref{eq4a} is questionable. Experiments (e.g. \cite{25,26,27}) were not able to yield clear-cut results on the validity of Eqs.~\eqref{eq1}-\eqref{eq4a}, and how to improve this simple approach: critical droplets are rare phenomena, typically one observes only the combined effect of nucleation and growth; also the results are often ``contaminated'' by heterogeneous nucleation events due to ions, dust, etc. \cite{28,29,30,31,32}, and since $j$ varies rapidly with the supersaturation, only a small window of parameters is suitable for investigation. Therefore this problem has been very attractive, in principle, for the study via computer simulation. However, despite numerous attempts (e.g.~\cite{8,9,10,16,21,22,23,24,33}), this approach is also hampered by two principal difficulties:
\begin{itemize}
	\item[(i)] Computer simulations can often only study a small number of decades in time, \cite{9,33}, which in typical cases correspond to small barriers $(\Delta F ^* \leq 10 k_BT)$ rather than the larger ones which are of more interest in the context of experiments.
	\item[(ii)] On the atomistic scale, it is a difficult and not generally solved problem to decide which particles belong to a droplet and which particles belong to its environment; the vapor-liquid interface is diffuse and fluctuating \cite{34,35}.
\end{itemize}
For these reasons, many of the available simulation studies have addressed nucleation in the simplistic Ising (lattice gas) model, \cite{9,10,18,22,33,35,36,37,38,39,40,41,42,43,44,45,46,47,48,49,50,51,52,53,54,55,56}, first of all since it can be very efficiently simulated, and secondly because one can define more precisely what is meant by a ``cluster''. Associating Ising spins $\sigma_i=+1$ at a lattice site $i$ with a particle, $\sigma _i=-1$ with a hole, originally ``clusters'' were defined as groups of up-spins such that each up-spin in a cluster has at least one up-spin as nearest neighbor belonging to the same cluster \cite{33}. However, now it is well understood that these ``geometrical clusters'' in general do not have much physical significance \cite{57,58,59,60,61}: e.g., it is known that there exists a line of percolation transitions, where a geometrical cluster of infinite size appears, in the phase diagram \cite{57}. This percolation transition is irrelevant for statistical thermodynamics of the model \cite{62,63,64,65}.

Based on the work of Fortuin and Kasteleyn \cite{66,67} on a correlated bond-percolation model, it is now understood that physically relevant clusters in the Ising model should not simply be defined in terms of spins having the same orientation and are connected by nearest neighbor bonds, as is the case in the ``geometrical clusters'', but in addition one has to require the bonds to be ``active'': bonds are ``active'' with probability $p$
\begin{equation}\label{eq4}
p(T) = 1 - \exp(-2J/k_BT) \;,
\end{equation}
$J$ being the Ising model exchange constant.

Due to Eq.~\eqref{eq4}, the ``physical clusters'' defined in this way are typically smaller than the geometrical clusters, and their percolation point can be shown to coincide with the critical point \cite{59,60,61}. A geometrical cluster hence can contain several physical clusters. Note that to apply Eq.~\eqref{eq4}, random numbers are used, and hence physical clusters are not deterministically defined from the spin configuration, but rather have some stochastic character. This presents a slight difficulty in using physical clusters in the study of cluster dynamics.

While Eq.~\eqref{eq4} has been used in the context of simulations of critical phenomena in the Ising model, applying very efficient Swendsen-Wang \cite{60} and Wolff \cite{68} simulation algorithms, this result has almost always been ignored in the context of simulations of nucleation phenomena \cite{18,22,47,48,49,50,51,52}. While it is allright to ignore the difference between geometrical and physical clusters in the limit $T \rightarrow 0$ (obviously $p(T) \rightarrow 1$ then, all bonds becoming active), this is completely inappropriate at higher temperatures.

The present work hence reconsiders this problem, studying both dynamical aspects of nucleation in the framework of the kinetic Ising model \cite{69,70} (without conservation laws), and the static properties of large droplets, applying the definition of ``physical clusters'' based on Eq.~\eqref{eq4} throughout. For comparison, we shall also occasionally use the ``geometrical'' cluster definition, to demonstrate that misleading conclusions would actually result in practice, for the temperatures that are commonly studied. The study will be generalized to Ising systems with free surfaces, where a boundary field $H_1$ acts \cite{55,56}. First of all, in this way also a systematic investigation of heterogeneous nucleation at planar walls becomes feasible; secondly, due to the reduction of the barrier $\Delta F^*_\text{het}$ in comparison to $\Delta F_\text{hom}^*$; nucleation for reasonably large values of $R^*$ becomes accessible to study.

In Sec.~\ref{sec2}, we consider the equilibrium of the lattice gas model for $\rho_v<\rho < \rho_\ell$ in systems in a $L \times L \times L$ geometry with periodic boundary conditions, to show that physical clusters do occupy precisely the volume predicted by the lever rule analysis, \cite{21,23,24,55} as they should when the thermodynamic limit is approached. We present evidence that physical clusters are correctly identified by both the lever rule method and the approach based on the ``atomistic'' identification of clusters based on Eq.~\eqref{eq4} at all temperatures, from zero temperature up to the critical temperature $T_c$. In contrast, Eq.\eqref{eq3}, which implies a spherical droplet shape, is found to work only at temperatures distinctly above the interface roughening transition temperature $T_R$ \cite{71,72}, even for very large radii $R$. We attribute these discrepancies to the fact that due to the anisotropy of the interface tension for our lattice model pronounced deviations of the average droplet shape from a sphere occur \cite{73,74,75,76}, presenting data on the shape of large droplets. In Sec.~\ref{sec3}, we describe our results on the dynamics of the droplet size distribution and on the attempt to find $R^*$ from the size where growth and shrinking processes of clusters are balanced. This study is also carried out for systems with a free surface, for which the contact angles for various values of the surface field have been estimated previously \cite{55,56}, since in this case much lower barriers (for large clusters) result, which is crucial for making this study feasible with manageable effort. However, the radii $R^*$ predicted from this analysis of kinetics show slight deviations from the radii $R^*$ predicted from $\Delta F^*_\text{het}$. Possible reasons for this discrepancy will be discussed. Finally, Sec.~\ref{sec4} summarizes our conclusions.

\section{Microscopically defined clusters versus macroscopic domains in thermal equilibrium}
\label{sec2}

As is obvious from Eq.~\eqref{eq2} and the reasoning behind it, this approach is adequate when one deals with the description of macroscopically large domains in equilibrium with a surrounding bulk phase. However, one needs to find an extension of the concept that can be applied also to nanoscopically small droplets, ``clusters'' in the lattice gas model that contain perhaps only of the order of 100 fluid particles. In this section, we want to confirm the idea that one must use the concept of ``physical clusters'' based on Eq.~\eqref{eq4} for this purpose, rather than the ``geometrical clusters'' that are so widely used when the lattice gas model is used to test nucleation theory concepts. While the geometrical clusters are appropriate if one works at extremely low temperatures where the clusters basically have the shape of small cubes \cite{52}, this region clearly is inappropriate when one has the application for vapor-to-liquid nucleation in mind, where droplets are spherical, and their interfaces are rough and fluctuating rather than smooth planar facets. In fact, many studies of nucleation in the lattice gas model have been made in $d=3$ dimensions at temperatures near $T/T_c = 0.6$ or thereabout; given the fact that the interfacial roughening transition of the Ising model on the simple cubic lattice is known to occur at about \cite{77,94} $T_R/T_c \approx 0.544$, i.e. (note $k_BT_c/J=4.51154$ \cite{77'}) $k_BT_R/J \approx 2.44$, it is clear that temperatures much closer to $T_c$ must be studied to render the assumption of a spherical droplet shape accurate. In fact, this assumption of a spherical droplet shape is accurate when the difference between the interfacial stiffness \cite{78} and the interfacial free energy becomes negligibly small. Numerical studies of Hasenbusch and Pinn \cite{79} indicate that this is only the case for $k_BT/J \geq 3.9$. As a consequence, it is clear that most of the existing studies of nucleation phenomena in the Ising model, that were based on geometrically defined clusters, and had to be done at much lower temperatures, are inconclusive: the deviation of the average droplet shape from a sphere enhances the surface term in Eq.~\eqref{eq2}; but the fluctuation corrections discovered for small droplets by the ``lever rule method'' \cite{21,23,24,55} show that $\gamma_{v\ell}(R) < \gamma_{v\ell}(\infty)$ for small $R$ and hence the surface term in Eq.~\eqref{eq2} is decreased. Thus, it hardly can be a surprise that some of the studies concluded that the nucleation barriers predicted by classical nucleation theory and the capillarity approximation (that ignores the $R$-dependence of $\gamma_{v\ell} (R))$ are too high, and others concluded they are too low, or even reported good agreement. We take the latter finding as indication that the two opposing effects have accidentally more or less canceled each other.

This problem is the motivation for the present section, which attempts to show that ``physical clusters'' based on Eq.~\eqref{eq4} are appropriate to identify clusters in the lattice gas model, irrespective of temperature and cluster size, and are equivalent to the droplets of the ``lever rule method'', for large enough droplets.

	\begin{figure}[ht]
	\begin{center}
	\includegraphics[clip=true, trim=12mm 20mm 5mm 10mm, angle=-90,width=0.8 \textwidth]{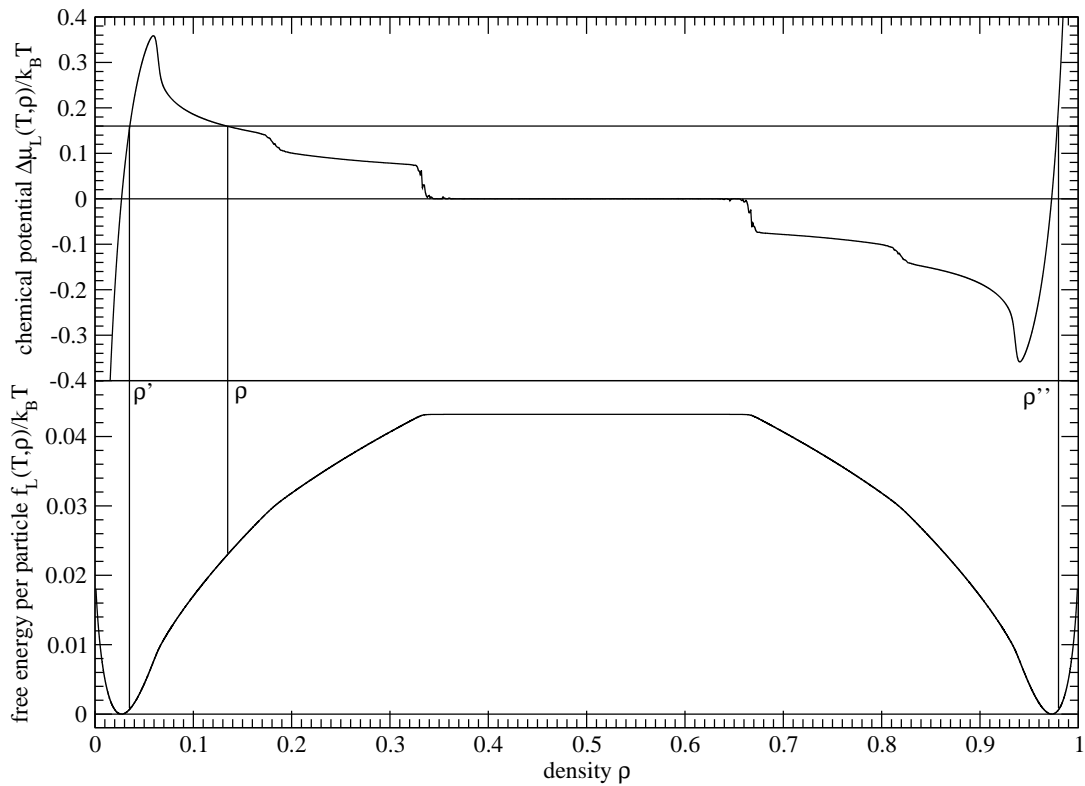}
	\caption{Illustration of the numerical procedure for determining the density triplets $\rho', \rho$ and $\rho''$ and the associated values $f_L(T, \rho')$, $f_L(T,\rho)$ and $f_L(T,\rho'')$ for a given choice of $T$ and $L$, according to the ``lever rule method''. The density $\rho$ must be chosen such that it is not too close to the peak of the curve $\Delta \mu _L/k_BT$ versus $\rho$, which is due to the ``droplet evaporation-condensation transition'', because all microstates that are sampled over should contain a droplet in the system. It must also be chosen not too close to the first kink in the curve $\Delta \mu_L/k_BT$ versus $\rho$, which is due to the transition of the droplet to a cylindrical shape (stabilized by the periodic boundary condition in the direction of the cylinders). Apart from these constraints, the choice of $\rho$ is arbitrary, and studying different choices of $\rho$ provides useful consistency checks on the results. The densities $\rho'$ and $\rho''$ then can be read off from the curve $\Delta \mu_L/k_BT$ versus $\rho$, since it is required that $\Delta \mu_L(T,\rho')=\Delta \mu_L(T,\rho)= \Delta \mu _L(T,\rho'')$. The corresponding values of the free energy densities then can be read off from the lower part of the figure, and using Eqs.~\eqref{eq7}, \eqref{eq8} with $N_\text{exc}=0$, the surface free energy $F_S$ is extracted. The actual data shown here refer to the case $k_BT/J=3.0$ and $L=20$.}
	\label{fig1}
	\end{center}
	\end{figure}

Fig.~\ref{fig1} recalls this approach: one samples for a system of volume $V= L \times L \times L$ the effective thermodynamic potential $f_L(T,\rho)$ per lattice site as a function of density $\rho$,
\begin{equation}\label{eq5}
f_L(T,\rho) = [F(N,V,T) - F(N=V\rho_{\ell}, V,T)]/V\,,
\end{equation}
where $N$ is the number of occupied lattice sites ($\rho_i=(1+\sigma_i)/2=+1$ with $\sigma_i=\pm 1$ the spin variable at lattice site $i$). Since phase coexistence between bulk liquid (at density $\rho_\ell$) and vapor (at density $\rho_v$) occurs at a chemical potential $\mu_\text{coex}$ that corresponds to the ``field'' (in magnetic notation) $H=0$, $\rho_\ell$ and $\rho_v$ are simply related to the spontaneous magnetization of the Ising ferromagnet as $\rho_\ell = (1+m_{sp})/2$, $\rho_v=(1-m_{sp})/2$, and $H$ translates into $\mu$ via $H=(\mu-\mu_\text{coex})/2=\Delta\mu/2$. The accurate sampling of $f_L(T,\rho)$ for large $L$ is a nontrivial task, it requires the use of advanced methods such as ``multicanonical Monte Carlo'' \cite{90} or ``Wang Landau sampling'' \cite{91,92} or ``successive umbrella sampling'' \cite{87,88}, see \cite{80,81} for background on such techniques. From Eq.~\eqref{eq5}, one defines a chemical potential function as a derivative,
\begin{align}\label{eq6}
\tilde{\mu} (N,V,T)&=\left.\frac{\partial F(N,V,T)}{\partial N}\right|_{V,T}\; ,\\
\label{eq6a}
\Delta \mu_L(T,\rho)&= \tilde{\mu}(N,V,T)-\mu_\text{coex}\;.
\end{align}
One recognizes that the isotherms $\Delta \mu_L(T,\rho)$ vs.~$\rho$ exhibit a loop: the homogeneous vapor remains stable also for some region where $\mu > \mu_\text{coex}$, until a peak occurs, which indicates the ``droplet evaporation/condensation transition'' \cite{21,82,89}: in the first regime where $\Delta \mu _L(T,\rho)$ decreases with $\rho$, a (more or less spherical or cubical) droplet coexists with surrounding vapor. Here, we are not interested in the further transitions that one can recognize from this curve, where the droplet changes shape from spherical to cylindrical, or to a slab configuration, etc. \cite{21,23,24}. Instead, we emphasize the key idea of the ``lever rule method'': one can identify a range of choices for the chemical potential $\mu$ where three states of the finite system can exist in equilibrium with the same chemical potential, namely a homogeneous vapor at density $\rho'>\rho_v$, a homogeneous liquid at density $\rho'' > \rho _\ell$, and a state where two-phase coexistence between the droplet and surrounding vapor occurs. Since the vapor in this case exists at the same chemical potential as the pure vapor, it must be of the same physical nature as the state with density $\rho'$, and similarly, the liquid in the droplet can be identified with the liquid at $\rho''$. Making now use of the fact that for large enough systems a system can be suitably decomposed into independent subsystems, we write for the free energy, with $V''$ the volume taken by the droplet, $V'=V-V''$,
\begin{equation}\label{eq7}
Vf_L(T,\rho)= V'f_L(T,\rho')+V'' f_L(T,\rho'') +F_S\; ,
\end{equation}
where the free energy densities $f_L(T,\rho')$ and $f_L(T,\rho'')$ are explicitly known, and also $f_L(T,\rho)$ is known: thus, when the droplet volume $V''$ is known, the surface free energy $F_S$ of the droplet, which is defined via Eq.~\eqref{eq7}, is determined. A similar decomposition can readily be written down for the particle number,
\begin{equation}\label{eq8}
N = V \rho = V' \rho' +V'' \rho'' + N_\text{exc}\;,
\end{equation}
where we have allowed for an excess number $N_\text{exc}$ of particles, to be associated with the interface. If we consider the definition of an ``equimolar dividing surface'' \cite{34}, $N_\text{exc}=0$, and then reading off $\rho'$ and $\rho''$ from the construction in Fig.~\ref{fig1} we see that Eq.~\eqref{eq8} readily yields $V'$ and $V''$ for the considered density $\rho$, and via Eq.~\eqref{eq7} we can immediately extract $F_S$ from the data. Note that these arguments do not invoke the assumption that the dividing surface needs to be a sphere. If one makes the assumption, $V'' = 4 \pi R^3/3$, and then one can write also $F_S=4 \pi R^2 \gamma_{v \ell}(R)$. Thus, it is assumed that all interactions of particles inside the droplet (volume region $V''$)  with particles inside the vapor (volume region $V'$) are restricted to the interfacial region, and hence can be accounted for by their contribution to the surface free energy $F_S$.

	\begin{figure}[ht]
	\begin{center}
	(a)
	\includegraphics[clip=true, trim=0cm 12mm 8mm 0cm, angle=-90,width=0.46 \textwidth]{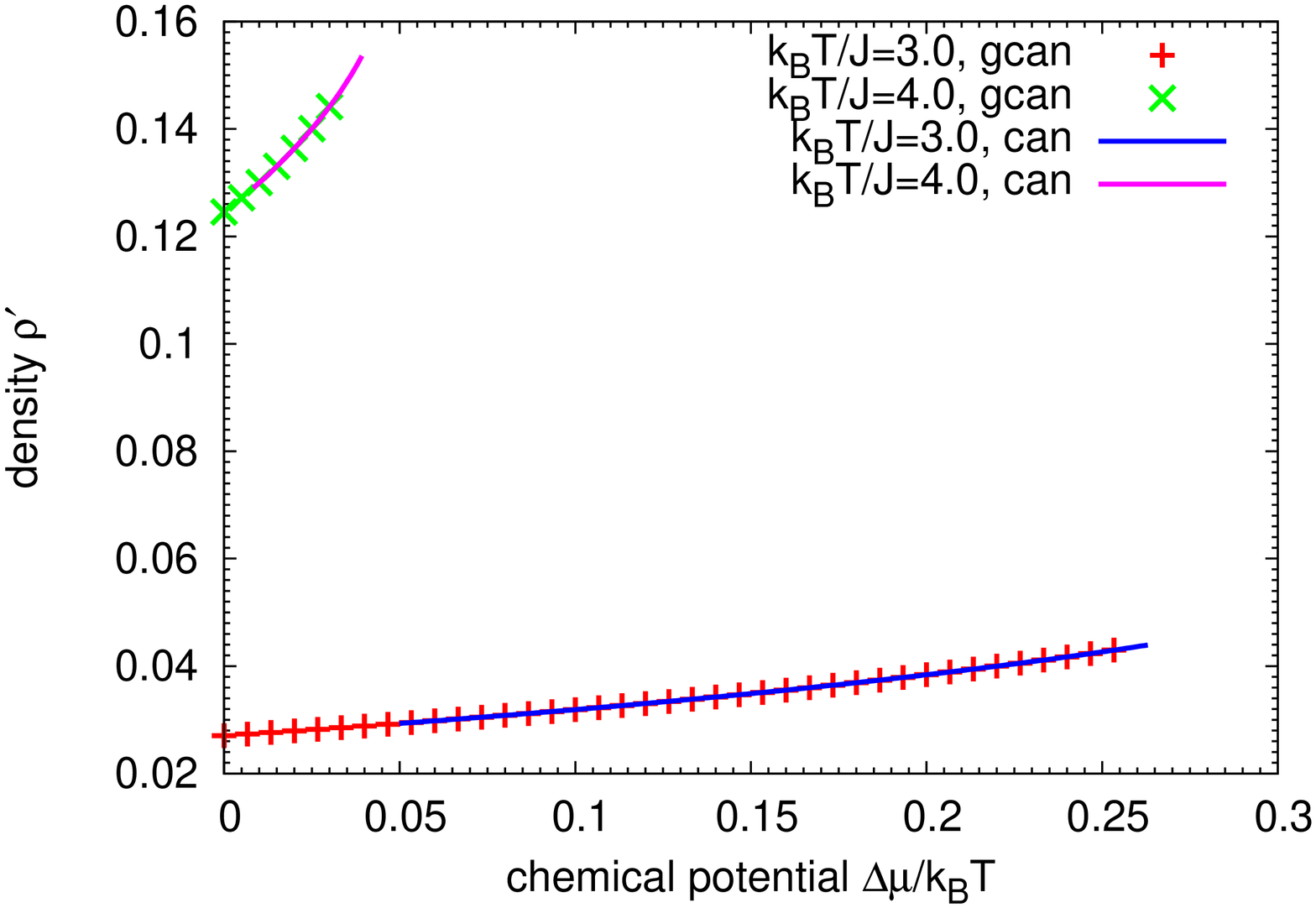}
	(b)
	\includegraphics[clip=true, trim=0cm 12mm 8mm 0cm, angle=-90,width=0.46 \textwidth]{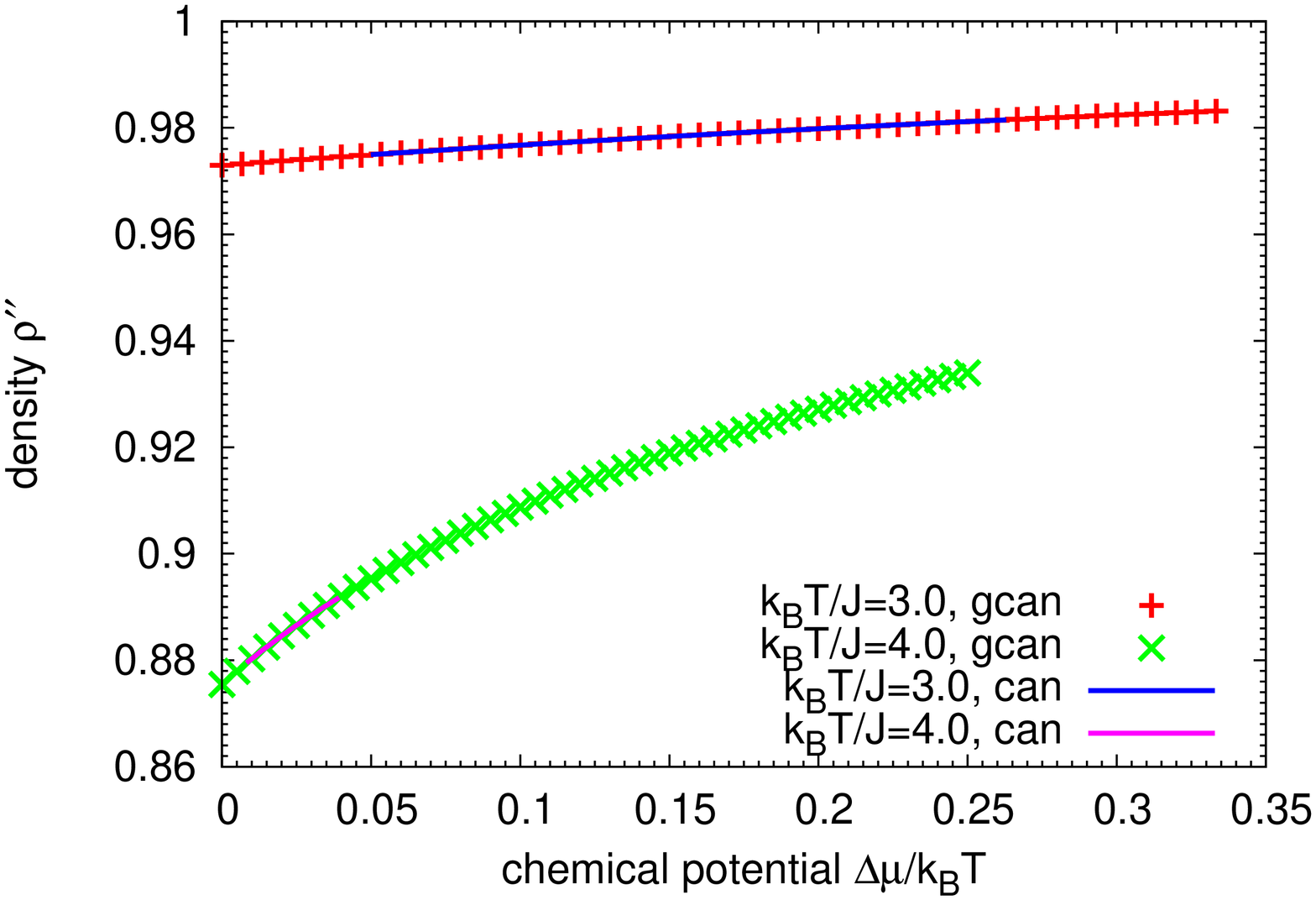}
	\caption{(Color online) (a) Plot of $\rho'(T,\Delta \mu)$ versus $\Delta \mu / k_BT$, for several temperatures $k_BT/J$ as indicated. The symbols represent data recorded in the grand-canonical ensemble of the lattice gas, while the curves are obtained in the canonical ensemble, recording $\Delta \mu = \Delta \mu (T,\rho)$ via the lattice version of the Widom particle insertion method \cite{55,84,85}. The chosen lattice size was $L=60$. (b) Same as (a), but for $\rho''(T,\Delta \mu)$. Both methods agree perfectly.}
	\label{fig2}
	\end{center}
	\end{figure}

However, the method defined via Eqs.~\eqref{eq7}, \eqref{eq8}, and illustrated in Fig.~\ref{fig1} becomes difficult to apply for very large droplets, because the sampling of $f_L(T,\rho)$ then becomes unreliable or would require an unaffordable effort. The method is also difficult to apply for small droplets, because then one must use relatively small simulation boxes to ensure the stability of the inhomogeneous state where a droplet coexists with surrounding vapor \cite{37}. Thus, it is very desirable to complement the approach by a more ``microscopic'' identification of droplets, and this is possible via Eq.~\eqref{eq4}. In particular, it has been shown that apart from finite size effects (see e.g. \cite{61}) that for $L \rightarrow \infty$ the spontaneous magnetization of the Ising ferromagnet $m_{sp}$ coincides with the percolation probability $P$, which is defined \cite{83} as the fraction of sites belonging to the largest ``physical cluster'' in the system. When we hence analyze a configuration at a density $\rho$, where (cf.~Fig.~\ref{fig1}) a large cluster is present in the system, using Eq.~\eqref{eq4} to define clusters the largest cluster will include $\ell$ sites, which we hence can associate with its (total) magnetization $M=m''V''$ where $m''$ is then the magnetization per site (note that the Ising magnet/lattice gas isomorphism implies that $\rho '' =(1+m'')/2$). As a consequence, we can obtain the droplet volume $V''$ from a ``measurement'' of the average size $\langle \ell \rangle$ of the largest cluster in the system via
\begin{equation}\label{eq9}
V''= \frac{\langle \ell \rangle}{m''} \; .
\end{equation}
Note that this volume in general differs from the volume of a geometrical cluster: If $\langle \ell_\text{geom} \rangle$ counts all occupied sites belonging to the geometrical cluster, and noting that the density in a large geometrical cluster is just the bulk density, namely $\rho''=(1+m'')/2$, the volume taken by the geometrical cluster is given by
	\begin{equation} \label{eq: Vgeom}
	V_\text{geom}''=  \frac{\langle \ell_\text{geom} \rangle}{\rho''} = \frac{2\langle \ell_\text{geom} \rangle}{1+m''}\;.
	\end{equation}
In the limit $L\rightarrow \infty$, where also $\langle \ell \rangle$ and $V''$ get macroscopically large, $m''$ tends to $m_{sp}$, while for finite $L$ it is clear that $m''$ slightly exceeds $m_{sp}$ (and $\rho''$ exceeds $\rho_\ell$, see Fig.~\ref{fig1}). However, recording the relation $\rho = \rho(T, \Delta \mu)$ [or the equivalent relation $m=m(T, H)$ of the Ising ferromagnet] very precisely is an easy task, see Fig.~\ref{fig2}, since both states at densities $\rho', \rho''$ (Fig.~\ref{fig1}) are homogeneous, not affected by heterophase fluctuations, and since the temperatures studied are still well below $T_c$, statistical fluctuations are small, and finite size effects are negligible. Fig.~\ref{fig2} presents representative results for both $\rho '(T,\Delta \mu)$ and $\rho''(T,\Delta\mu)$ versus $\Delta \mu$. Note we also have used the lattice version of the Widom particle insertion method \cite{55,84,85} to record the inverse function $\Delta \mu = \Delta \mu (T,\rho)$ from simulations in the canonical ensemble, where $\rho$ was chosen as the independent control variable. The perfect agreement between both approaches not only serves as a test of the accuracy and correctness of our numerical procedures, but also shows that for the chosen temperatures and linear dimensions finite size effects on states in ``pure'' phases are completely negligible, since finite size effects are known to differ in the two ensembles \cite{80,81}, but are not detected here at all.

	\begin{figure}[ht]
	\begin{center}
	(a)
	\includegraphics[clip=true, trim=0cm 12mm 0mm 0cm, angle=-90,width=0.46 \textwidth]{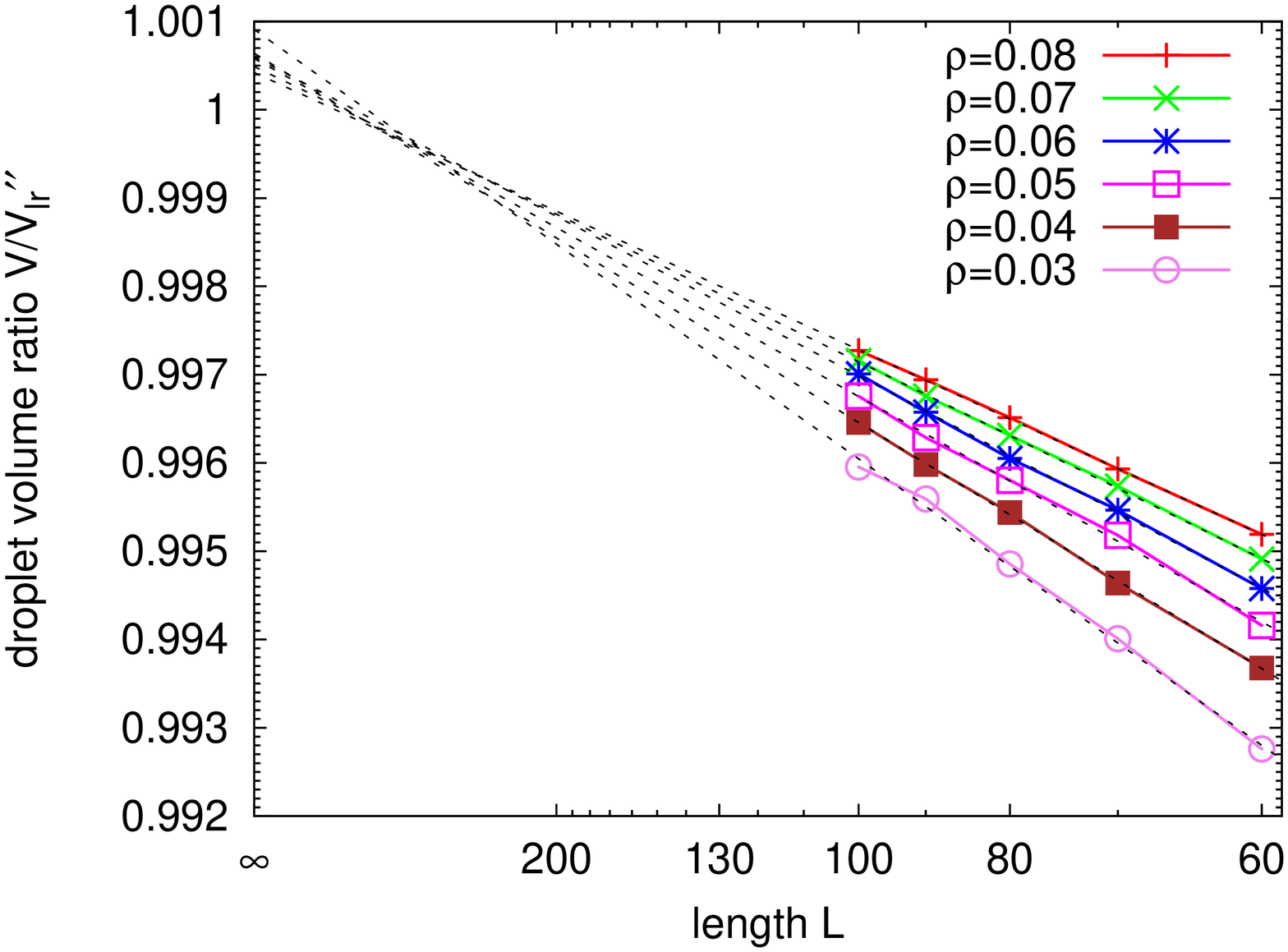}
	(b)
	\includegraphics[clip=true, trim=0cm 12mm 0mm 0cm, angle=-90,width=0.46 \textwidth]{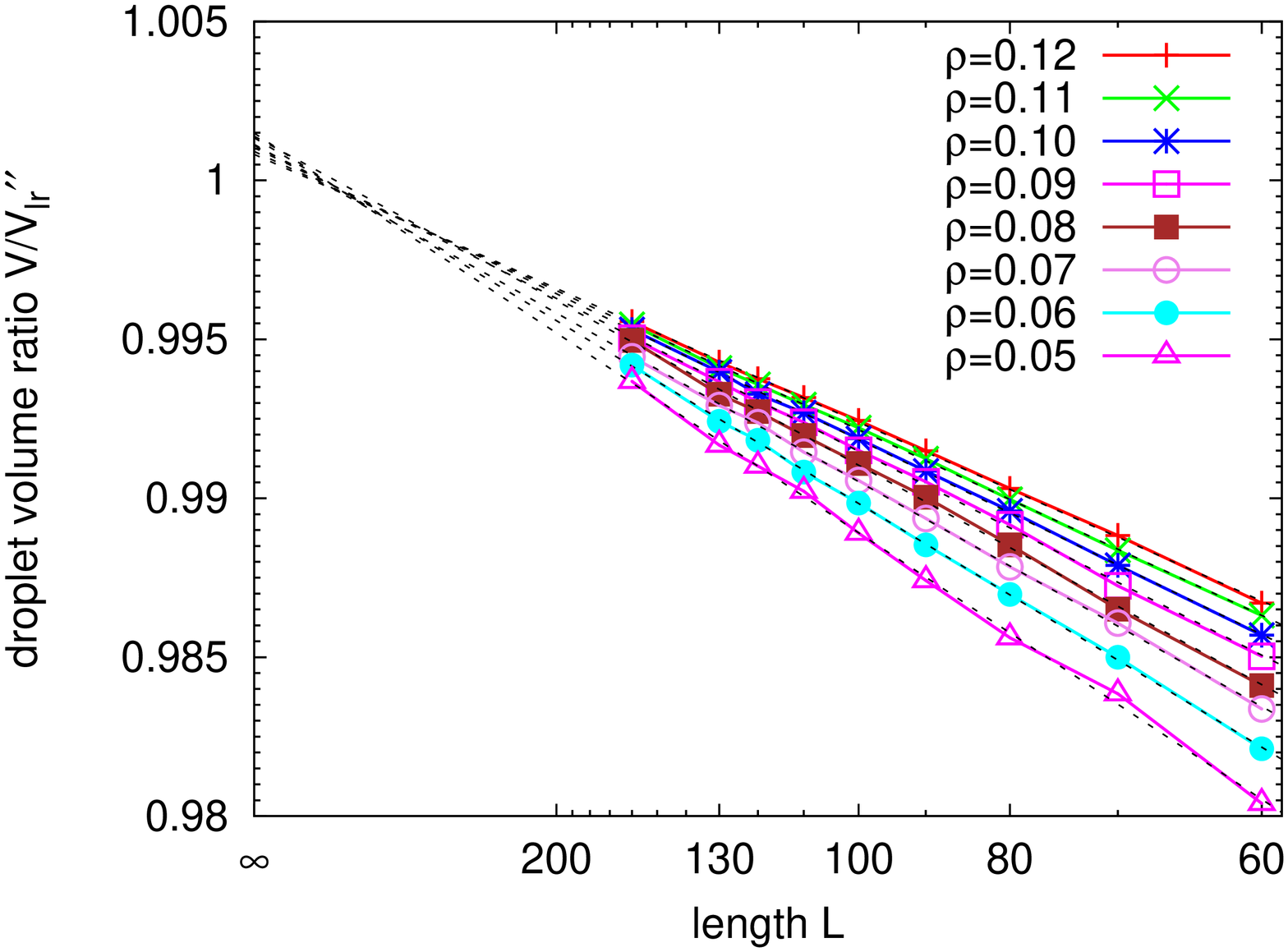}
	(c)
	\includegraphics[clip=true, trim=0cm 12mm 0mm 0cm, angle=-90,width=0.46 \textwidth]{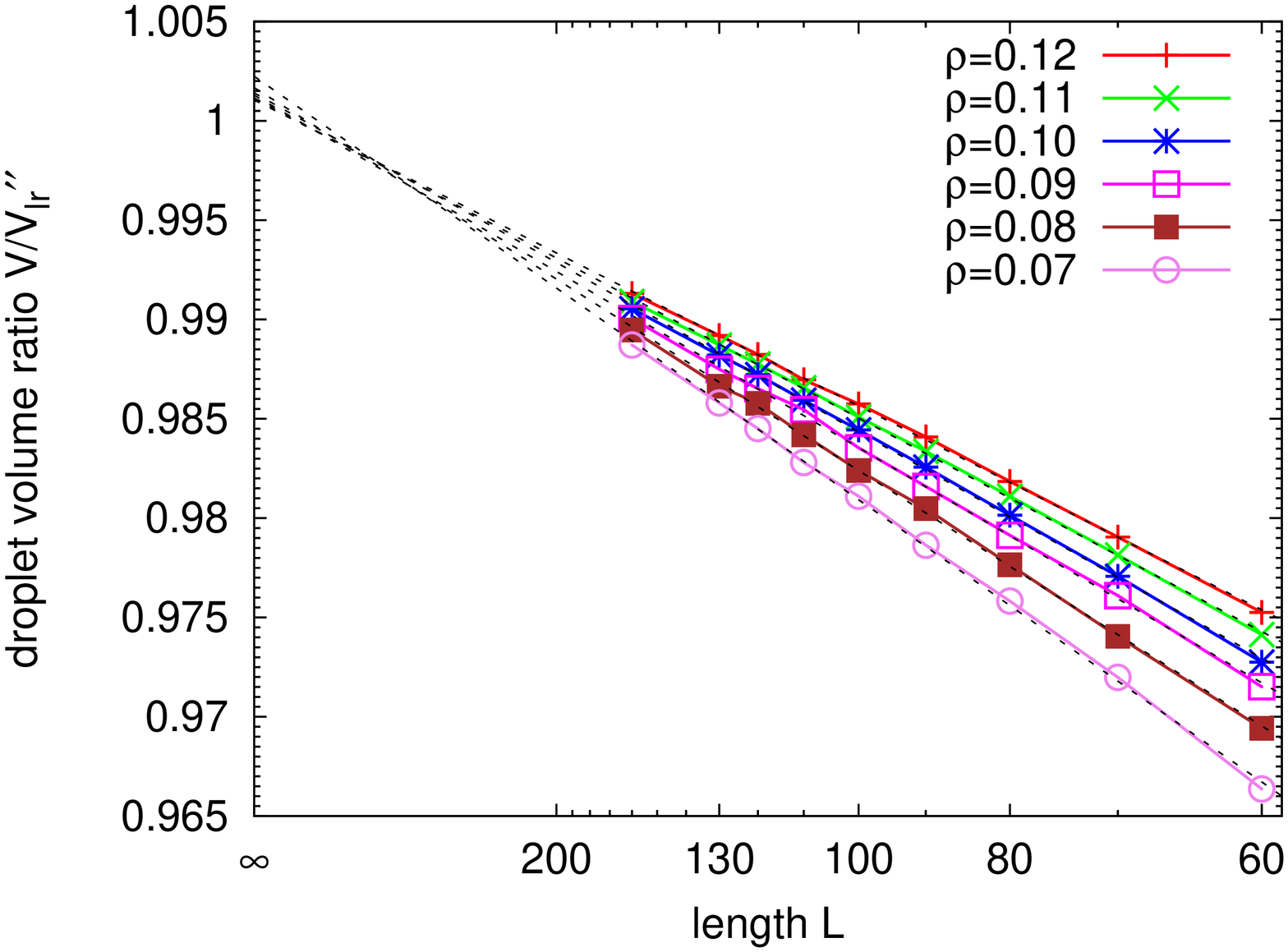}
	(d)
	\includegraphics[clip=true, trim=0cm 12mm 0mm 0cm, angle=-90,width=0.46 \textwidth]{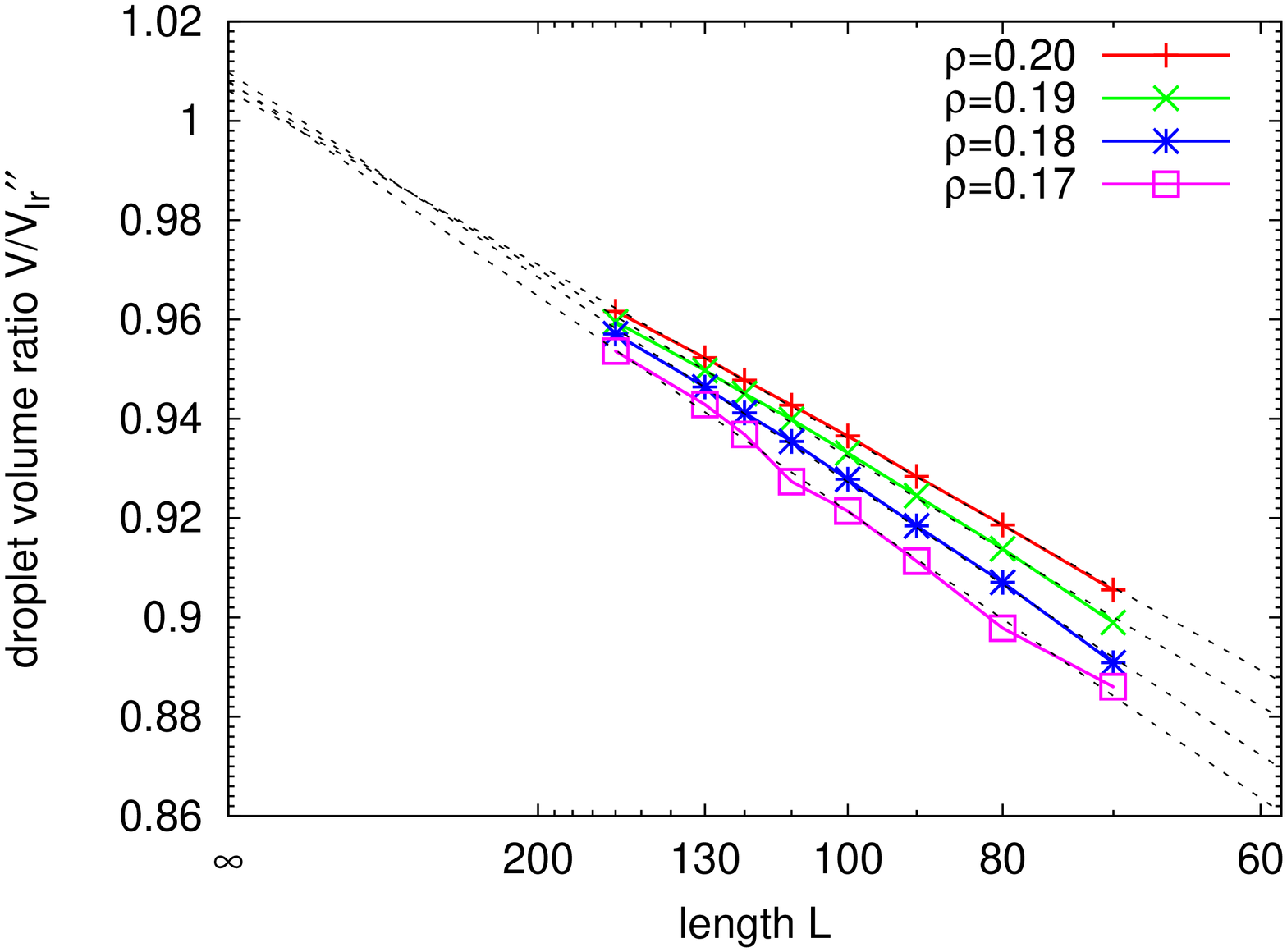}
	\caption{(Color online) Plot of the ratio $V''/V''_{lr}$ of the droplet volume as obtained from the ``Coniglio-Klein-Swendsen-Wang'' \cite{59,60} cluster definition, using Eq.~\eqref{eq4} to define physical clusters, and their volume $V''$ (Eq.~\eqref{eq9}), and from the lever rule (Eq.~\eqref{eq10}), as a function of $1/L$, for four temperatures: $k_BT/J=2.0$ (a), $2.6$ (b), $3.0$ (c) and $4.0$ (d). Various densities are included for each temperature, as indicated.}
	\label{fig3}
	\end{center}
	\end{figure}

To record $V''$ as defined in Eq.~\eqref{eq9} from the simulations, we performed simulations in the canonical ensemble, choosing various values of $\rho$, and equilibrate a large droplet coexisting with surrounding vapor. The initial state is then chosen putting a droplet with the size predicted by the lever rule (and density $\rho = (1+m_{sp})/2$) into the simulation box, which then is carefully equilibrated. The standard method to simulate the Ising model with conserved magnetization (which corresponds to the lattice gas model in the canonical $NVT$ ensemble) is the ``spin exchange algorithm'' \cite{80,81}. However, the standard nearest neighbor exchange method implies that any local excess of magnetization (or density, respectively) can only relax diffusively, and the resulting ``hydrodynamic slowing down'' \cite{80,81} hampers the fast approach towards thermal equilibrium. We thus used instead a single spin-flip algorithm in which the total magnetization is restricted to two neighboring values $\{M-2, M\}$: So if the system has magnetization $M$, flipping a down-spin (which would mean a transition $M\to M+2$) is automatically rejected, and if the system is in the state $M-2$, flipping of up-spins is forbidden. For large systems ($L\to \infty$) any corrections to a strictly canonical simulation at a magnetization per spin $m=M/L^3$ are of order $1/L^3$ and hence negligible.

\begin{figure}[ht]
	\begin{center}
	(a)
	\includegraphics[clip=true, trim=0cm 12mm 8mm 0cm, angle=-90,width=0.46 \textwidth]{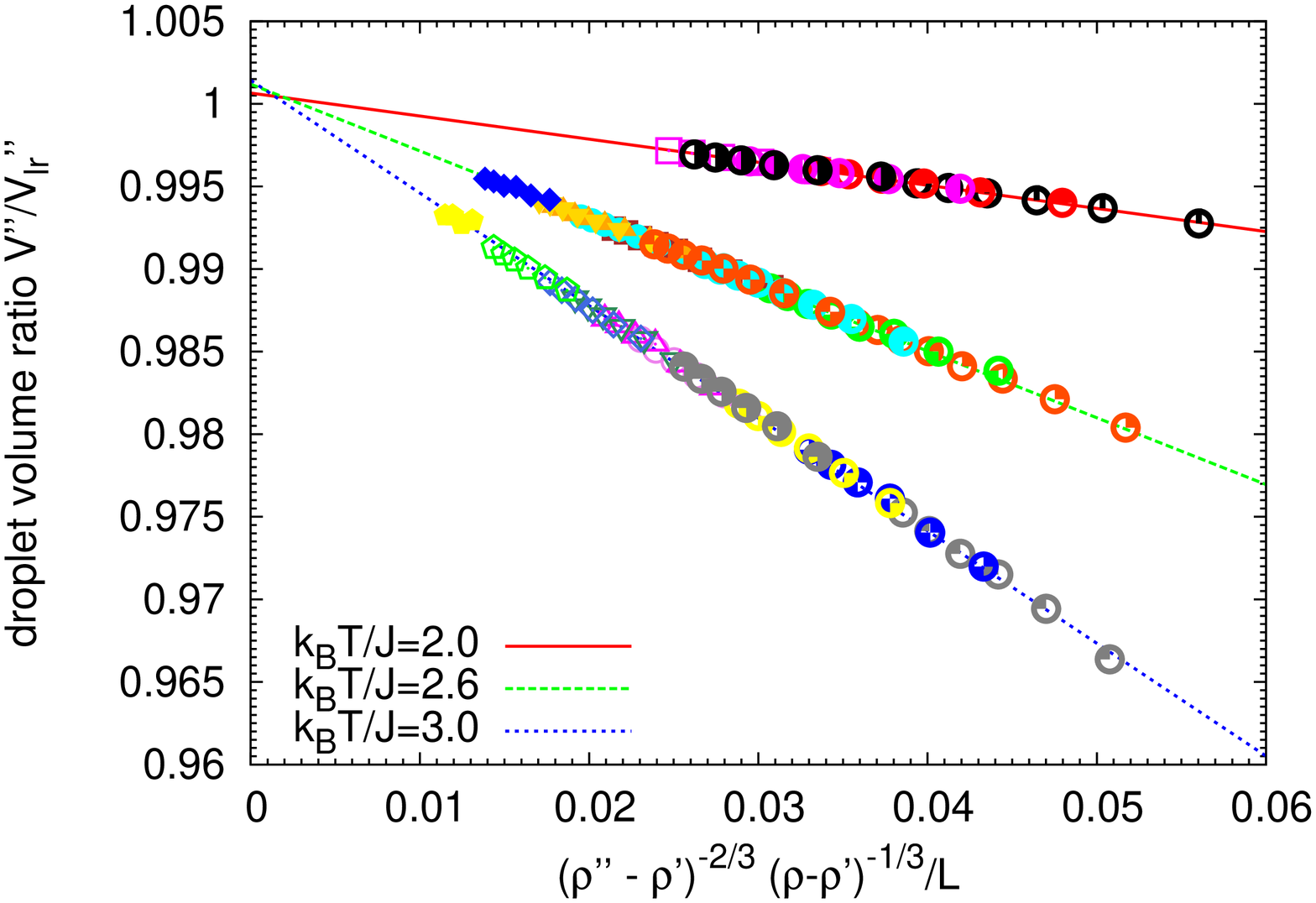}
	(b)
	\includegraphics[clip=true, trim=0cm 12mm 8mm 0cm, angle=-90,width=0.46 \textwidth]{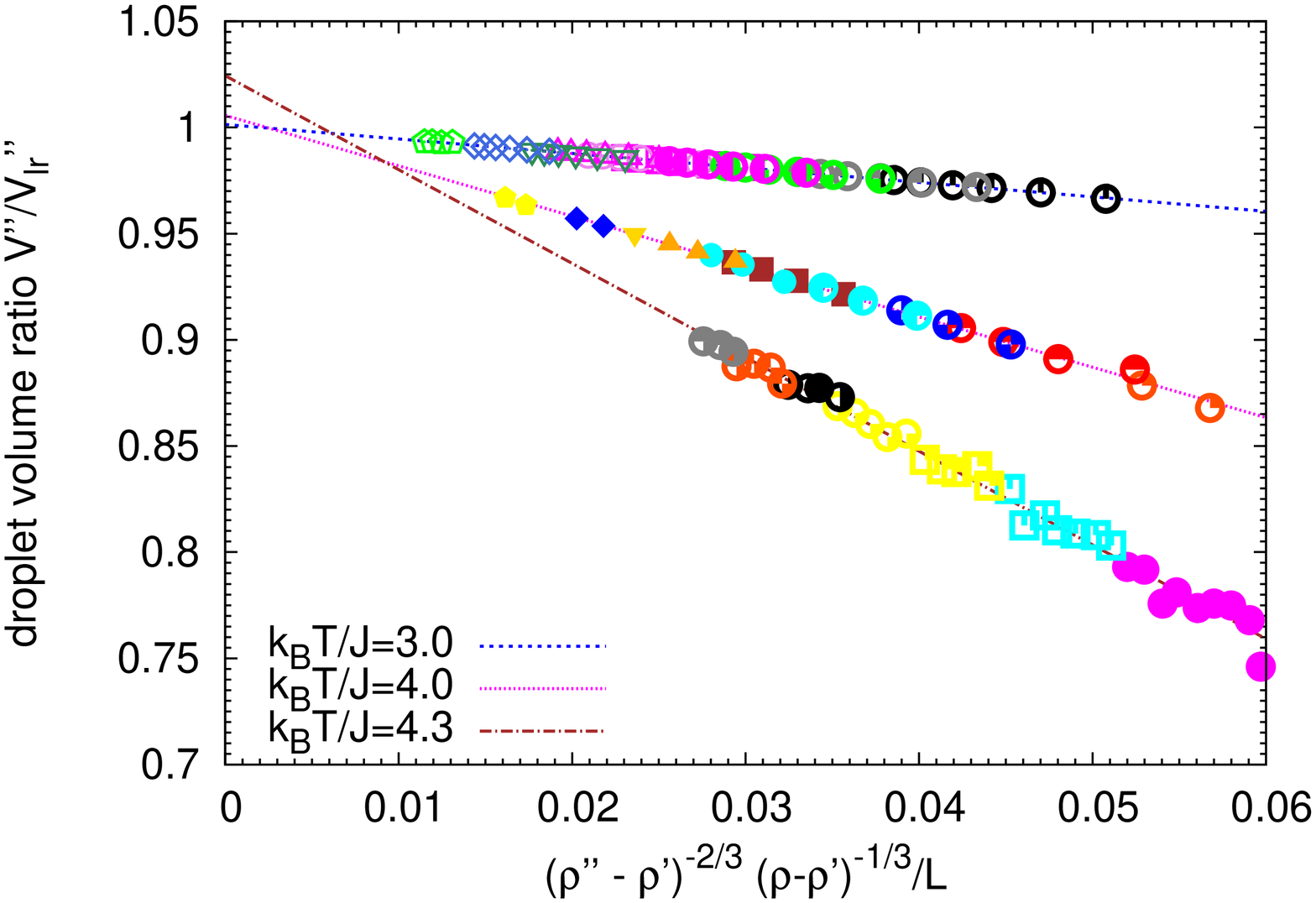}
	\caption{(Color online) Plot of $V''/V_{lr}''$ versus $(\rho''-\rho')^{-2/3}(\rho-\rho')^{-1/3}/L$ at low temperatures (a) and at temperatures closer to $T_c$ (b), including all densities $\rho$ that were analyzed. Straight line fits to the data are included. Note that the different symbols correspond to different choices of $\rho$ in Fig.~\ref{fig3}.}
	\label{fig4}
	\end{center}
	\end{figure}

Choosing very large systems (up to $L=160$, rather than $L=20$ as used in Fig.~\ref{fig1}) the ratio of $V''$ as found from Eq.~\eqref{eq9} and $V_{lr}''$ from the lever rule (with the assumption $N_\text{exc} \equiv 0)$, i.e.
\begin{equation}\label{eq10}
V_{lr}'' = V\frac{\rho - \rho'}{\rho''-\rho'} \; ,
\end{equation}
is plotted vs. $1/L$ in Fig.~\ref{fig3} for several temperatures and various densities $\rho$. These data show that $V''/V''_{lr}\rightarrow 1$ as $L \rightarrow \infty$, irrespective of temperature and density (in the density region where a droplet not affected by the periodic boundary conditions is present, as explained in Fig.~\ref{fig1}). The fact that $V''/V_{lr}''$ extrapolates to unity for $L\to \infty$ not precisely, but only within some error, is due to the fact that statistical errors affect both the estimation of $\langle \ell \rangle$ and of $V_{lr}''$ (via errors in the estimation of $\Delta \mu$ and hence $\rho'$). As expected, using the proper definition of physical clusters, one can work at arbitrary temperatures, both at $T < T_R$ (a), $T$ slightly above $T_R$ (b) or $T$ rather close to $T_c$ (d). While in cases (a) the simple geometrical cluster definition would also work, since essentially all bonds are ``active'' ($p(T)$ is almost unity \{Eq.~\eqref{eq4}\}), and the non-spherical shape of the clusters does not matter in this context. But the geometrical cluster definition would clearly break down in case (d) due to the proximity of the percolation transition that occurs for geometrical clusters at a density not much larger than those included in Fig.~\ref{fig3}(d) \cite{57,64,65}.
On the other hand, we note from the fact that there always occurs asymptotically in the relation $V''/V''_{lr}$ a correction of order $1/L$, that for physical clusters the assumption $N_\text{exc}=0$, that is often (but not always \cite{24}) made in the lever rule method, does not hold: i.e., when we assume that $N_\text{exc}$ is proportional to the surface area of the droplet, we can write
\begin{equation}\label{eq11}
N_\text{exc}=C(V'')^{2/3} \rho_\text{exc}\quad ,
\end{equation}
where $C$ is a geometrical factor $(C=(36 \pi)^{1/3}$ for a spherical droplet, $C=6$ for a cube), and $\rho_\text{exc}$ is an excess density of the particles due to the interface of the droplet. Using $V_{lr}''$ from Eq.~\eqref{eq10} as a first-order estimate in Eq.~\eqref{eq11}, one readily finds that the term $N_\text{exc}/V$ in Eq.~\eqref{eq8} yields a $1/L$ correction,
\begin{equation}\label{eq12}
\frac{N_\text{exc}}{V} = \frac{C}{L} \left(\frac{\rho - \rho'}{\rho'' - \rho'}\right)^{2/3} \rho_\text{exc} \quad ,
\end{equation}
and hence Eq.~\eqref{eq8} would yield instead of Eq.~\eqref{eq10}
\begin{equation}\label{eq13}
\begin{split}
\frac{V_{lr}''(\text{corrected})}{V_{l r}''}
&=1- \frac{C}{L} \left(\frac{\rho - \rho '}{\rho'' - \rho'}\right) ^{2/3} \frac {\rho_\text{exc}}{\rho-\rho '}\quad , \quad L \rightarrow \infty \; \\
&=1- \frac{C\rho_\text{exc}}{L} (\rho'' - \rho')^{-2/3} (\rho- \rho')^{-1/3}
\end{split}
\end{equation}
which is qualitatively in accord with Fig.~\ref{fig3}. In order to test Eq.~\eqref{eq13} and obtain estimates for the temperature dependence of $\rho_\text{exc}$, Fig.~\ref{fig4} plots our numerical results for the ratios $V''/V_{lr}''$ versus $(\rho''-\rho')^{-2/3}(\rho-\rho')^{-1/3}/L$. We see a very good data collapse at straight lines going through unity at the ordinate within numerical error at all studied temperatures.



	\begin{figure}[ht]
	\begin{center}
	(a)
	\includegraphics[clip=true, trim=0cm 12mm 0mm 0cm, angle=-90,width=0.46 \textwidth]{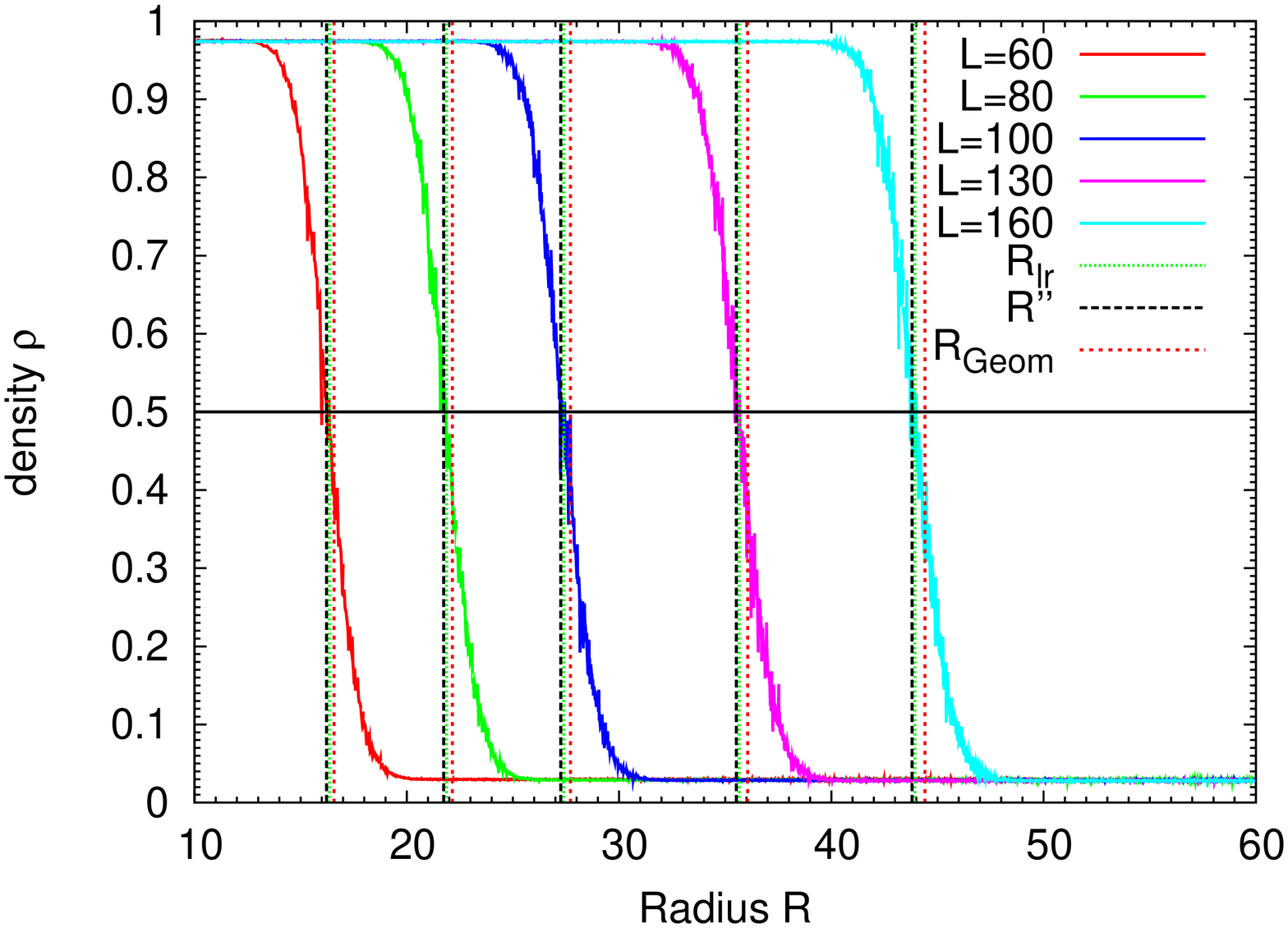}
	(b)
	\includegraphics[clip=true, trim=0cm 12mm 0mm 0cm, angle=-90,width=0.46 \textwidth]{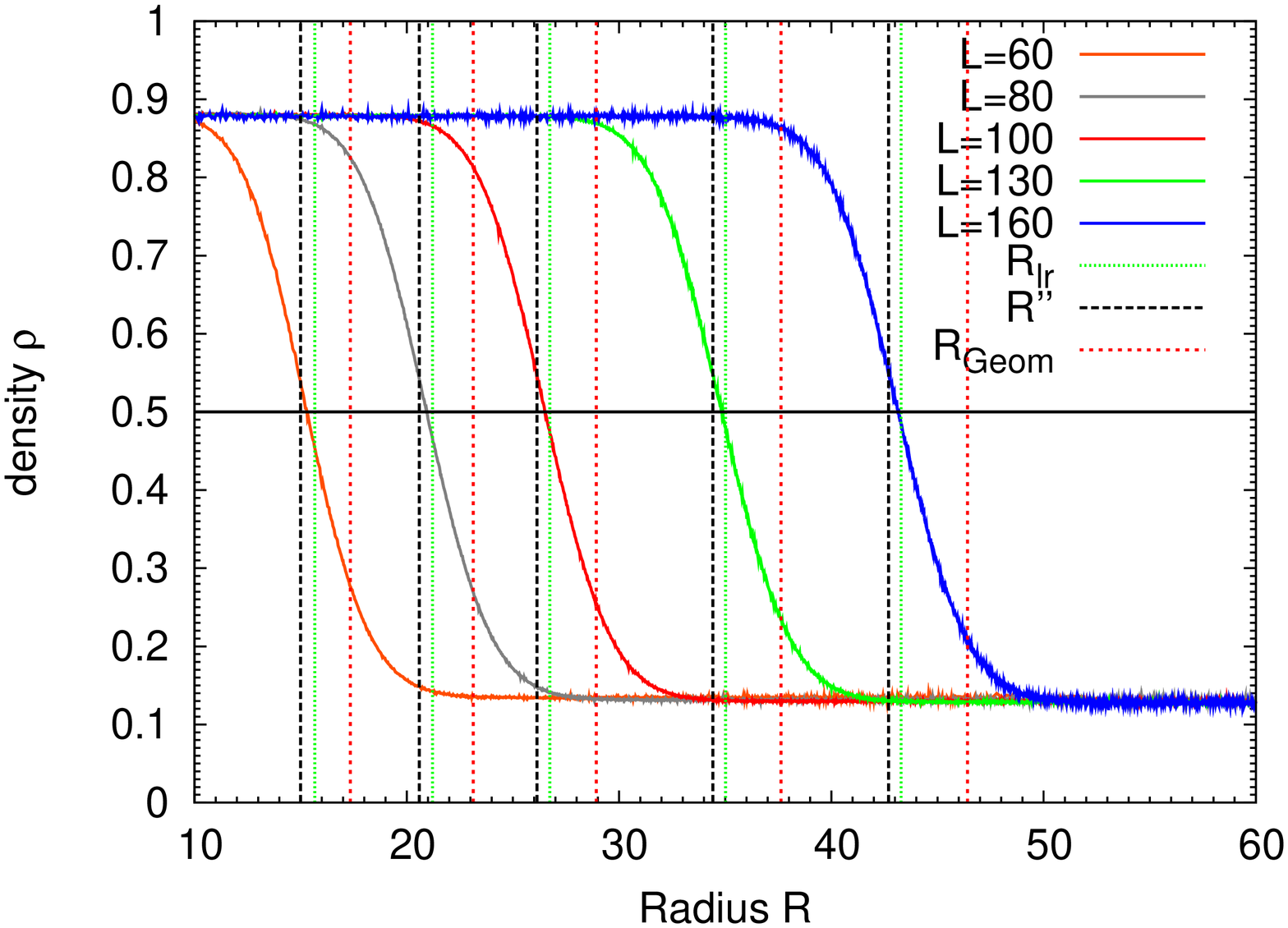}
	(c)
	\includegraphics[clip=true, trim=0cm 12mm 0mm 0cm, angle=-90,width=0.46 \textwidth]{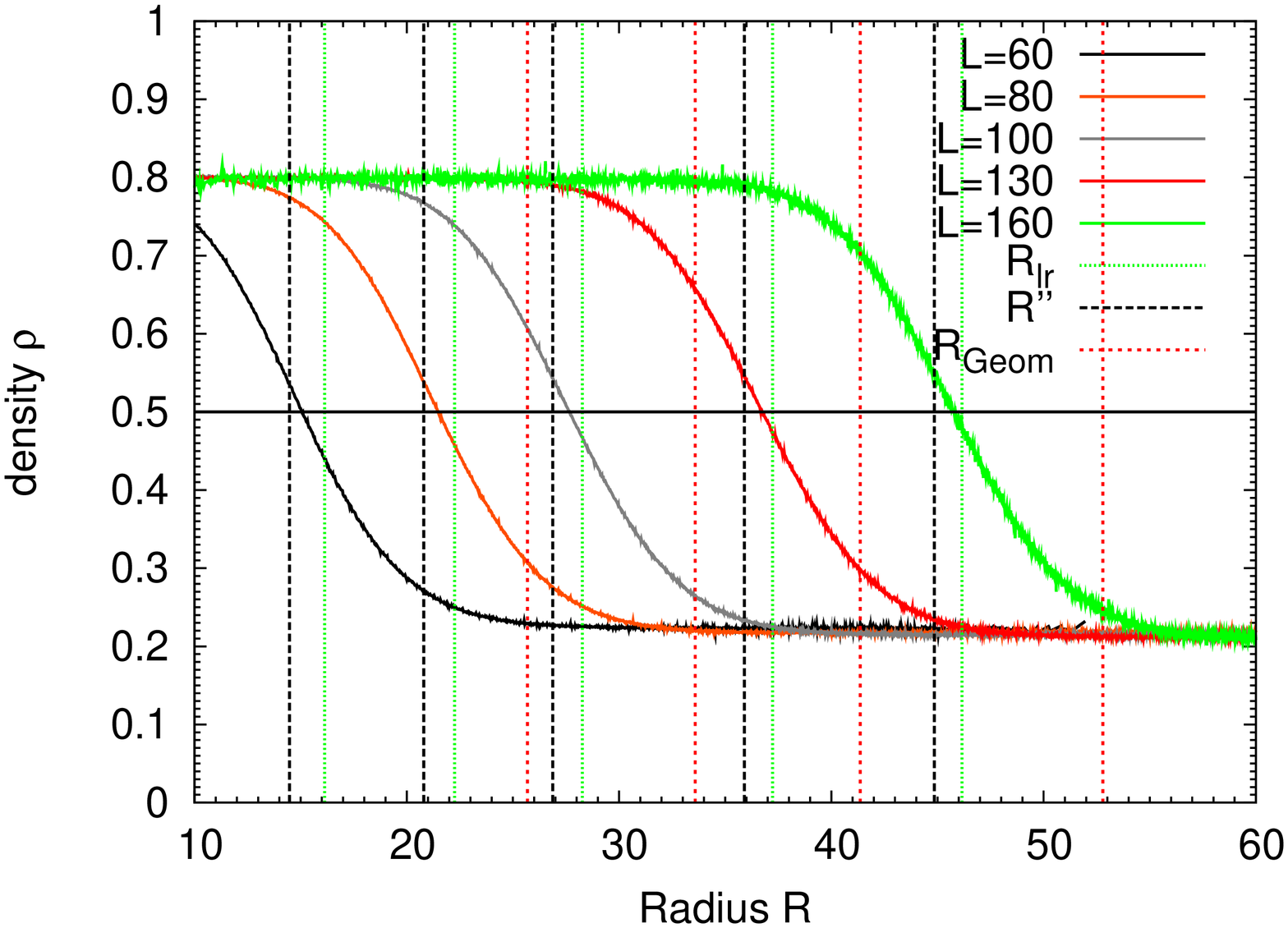}
	(d)
	\includegraphics[clip=true, trim=0mm 5mm 0mm 0cm, angle=-90,width=0.46 \textwidth]{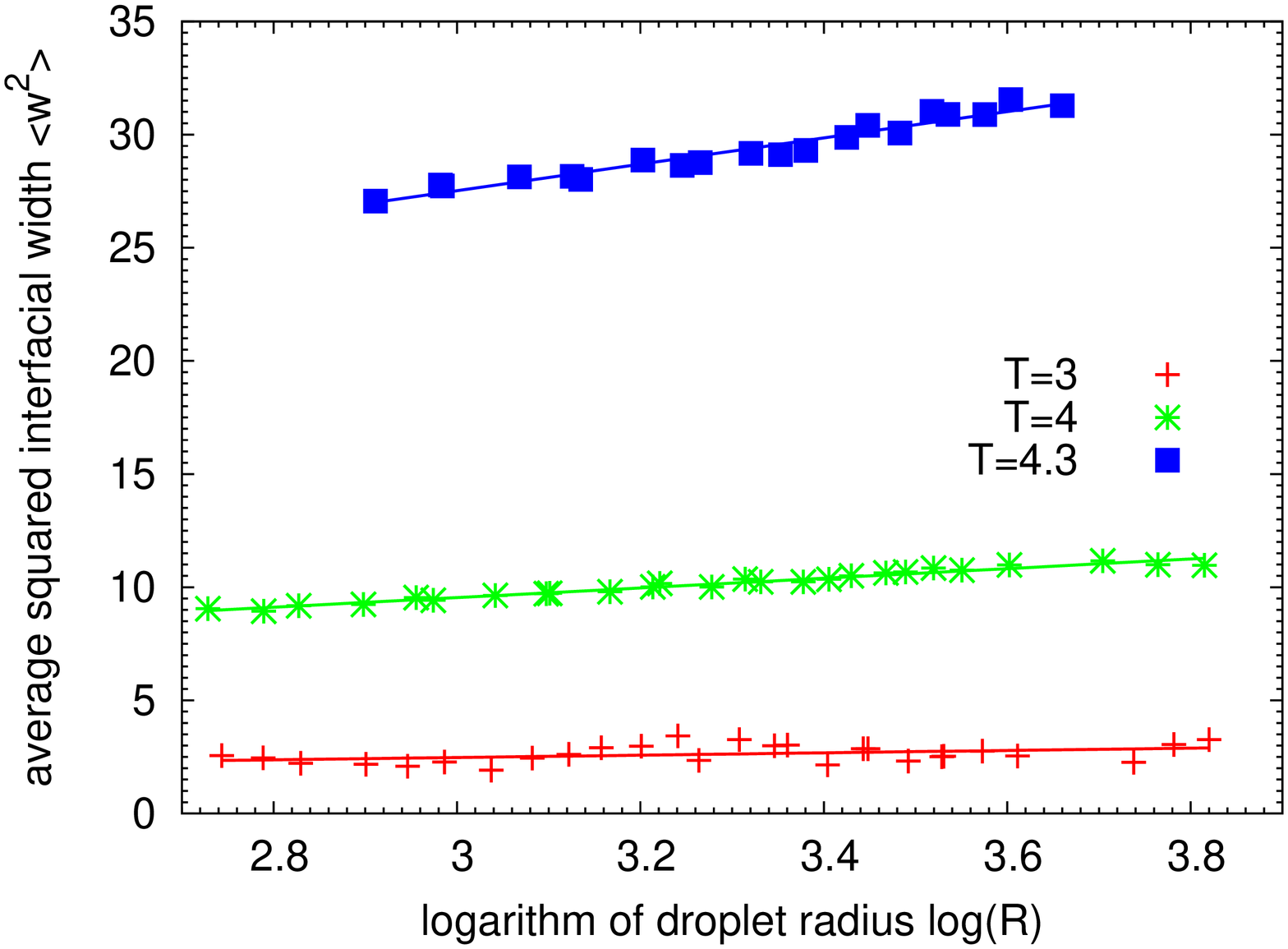}
	\caption{(Color online) Radial density profiles $\rho(R)$ plotted vs. $R$ at $k_BT/J=3.0$ (a), $T=4.0$ (b) and $k_BT/J=4.3$ (c), for a box of size $L\times L \times L$ with $L= 60, 80, 100, 130, 160$ (from left to right). The vapor density and the liquid density are independent of $L$, as expected. Vertical straight lines indicate the radii that follow (assuming $R=(3V''/4 \pi )^{1/3}$, i.e.~a spherical shape) from Eq.~\eqref{eq9}  with $V''= \langle \ell \rangle /m''$ or from the simple geometrical cluster definition, Eq.~\eqref{eq: Vgeom}, where $V_\text{geom}''= 2 \langle \ell_\text{geom}\rangle/(1+m'')$ or from the lever rule Eq.~\eqref{eq10}. A horizontal line at $\rho=0.5$ is also drawn, since for $R \rightarrow \infty$ the density profile should become symmetric with respect to this line, due to the spin reversal symmetry of the Ising model, and then the correct cluster definition must yield a cluster radius compatible with this inflection point of the profile. All data are obtained from averages over several hundred independent droplet configurations. The dotted vertical lines are the radii predicted from the lever rule. Part (d) gives a plot of the average squared interfacial width $\langle w^2 \rangle$ vs. $\ln R$ (in this plot, $R$ is taken from the condition $\rho(R)=0.5$).}
	\label{fig5}
	\end{center}
	\end{figure}

Since the data in Fig.~\ref{fig3} suggest that for ``physical droplets'' in the lattice gas model an appreciable ``interfacial adsorption'' (as expressed in $\rho_\text{exc} $ in Eq.~\eqref{eq11}) occurs, it is of interest to not only study the average volume of the droplets $\langle \ell \rangle$ but also their radial density profile (Fig.~\ref{fig5}). It is seen that for radii $R$ which are in the range from 10 to 20 lattice constants (corresponding to droplet volumes in the range from about 4500 to 36000, so these are already clusters of a mesoscopic size, with a huge nucleation barrier far beyond observation in a simulation of nucleation events or in experiments) the profiles are very broad, and their width increases slightly with increasing droplet size. For comparison, the prediction for the cluster radius resulting from the widely used standard geometrical definition of clusters is also included: it happens that this geometrical radius is still fairly close to the correct radius at $k_BT/J=3.0$. Of course, closer to $T_c$ the geometrical cluster definition yields completely unreasonable results, due to the onset of percolation phenomena, and for $k_BT/J=4.0$ and $4.3$, the geometrical radii are indeed unreasonably large.

It is interesting to note (Fig.~\ref{fig5}d) that the width of the interfacial profile increases with increasing $R$. This phenomenon is well-known for planar interfaces and attributed to capillary waves. Since for a large droplet the surface is locally planar, most of the capillary wave spectrum is not affected by the interface curvature. Thus we may, as a first approximation, take over the result for the broadening of a planar interface of linear dimension $L$, replacing $L$ by the droplet radius $R$ \cite{97,98,99}
	\begin{equation}
	\langle w^2\rangle = w_0^2 + \frac{1}{4\widetilde{\gamma}} \ln\left(\frac{R}{\lambda_\text{min}}\right)
	\end{equation}
where $w_0$ is the ``intrinsic width'' of the interfacial profile, $\widetilde{\gamma}$ the ``interfacial stiffness'' (note that a factor $1/k_BT$ is absorbed in its definition) and $\lambda_\text{min}$ a short wavelength cutoff, whose precise value is not known. Near $T_c$ the interfacial stiffness coincides with the interfacial free energy, while $\widetilde{\gamma}\to \infty$ at $T\to T_R$, and then the capillary wave broadening disappears. If one accepts the above formula, and uses the data of Hasenbusch and Pinn \cite{79}, one predicts for the slope $[4\widetilde{\gamma}]^{-1} \approx 2.5$ ($k_BT/J=4.0$) or $7.98$ ($k_BT/J=4.3$), respectively. The actually observed slopes of the $\ln R$ term are actually somewhat smaller, namely about 2.13 at $k_BT/J=4.0$ and $5.83$ at $k_BT/J=4.3$, but of the same order of magnitude.

\begin{figure}[t]
	\begin{center}
	(a)
	\includegraphics[clip=true, trim=0cm 12mm 0mm 0cm, angle=-90,width=0.46 \textwidth]{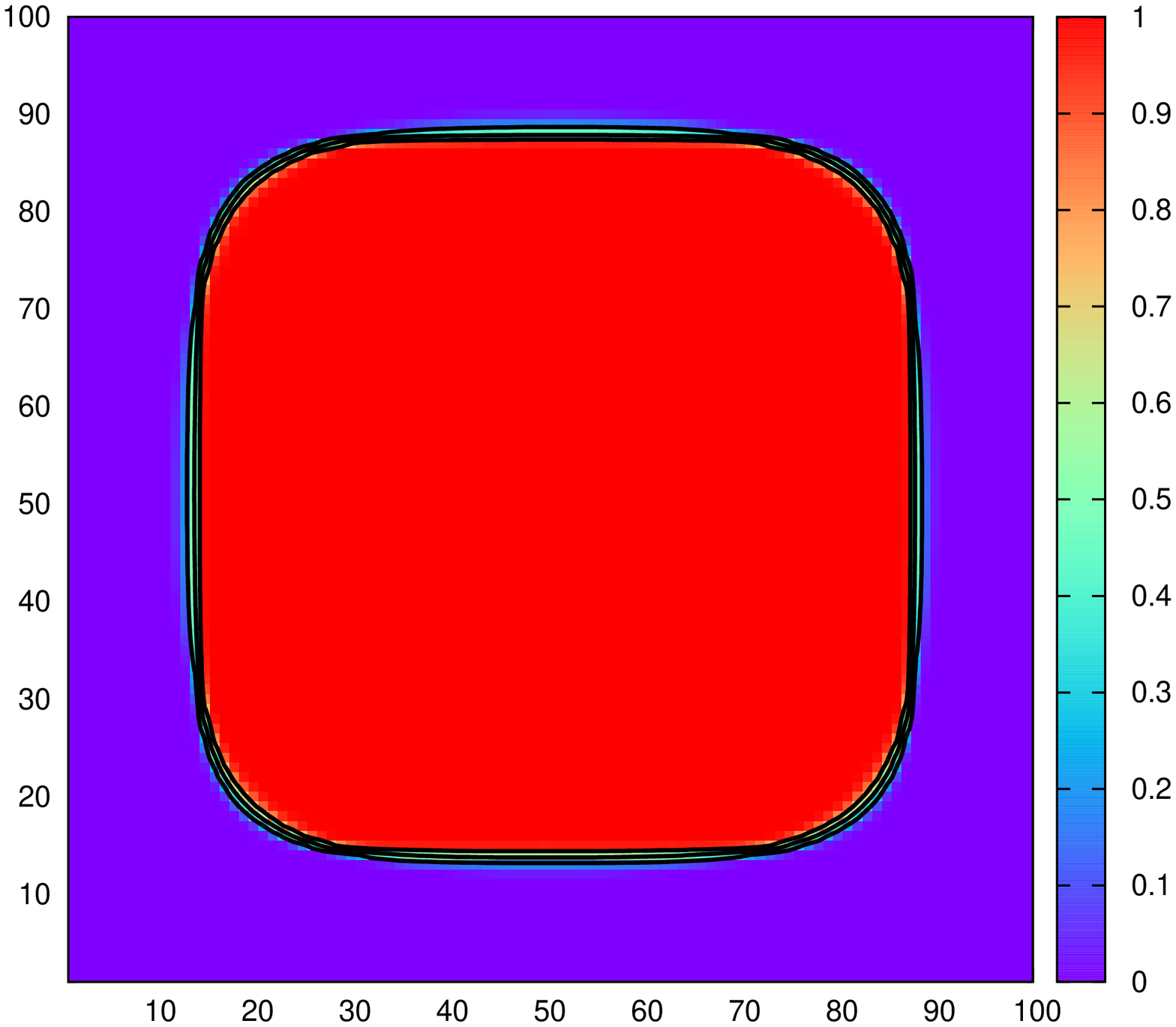}
	(b)
	\includegraphics[clip=true, trim=0cm 12mm 0mm 0cm, angle=-90,width=0.46 \textwidth]{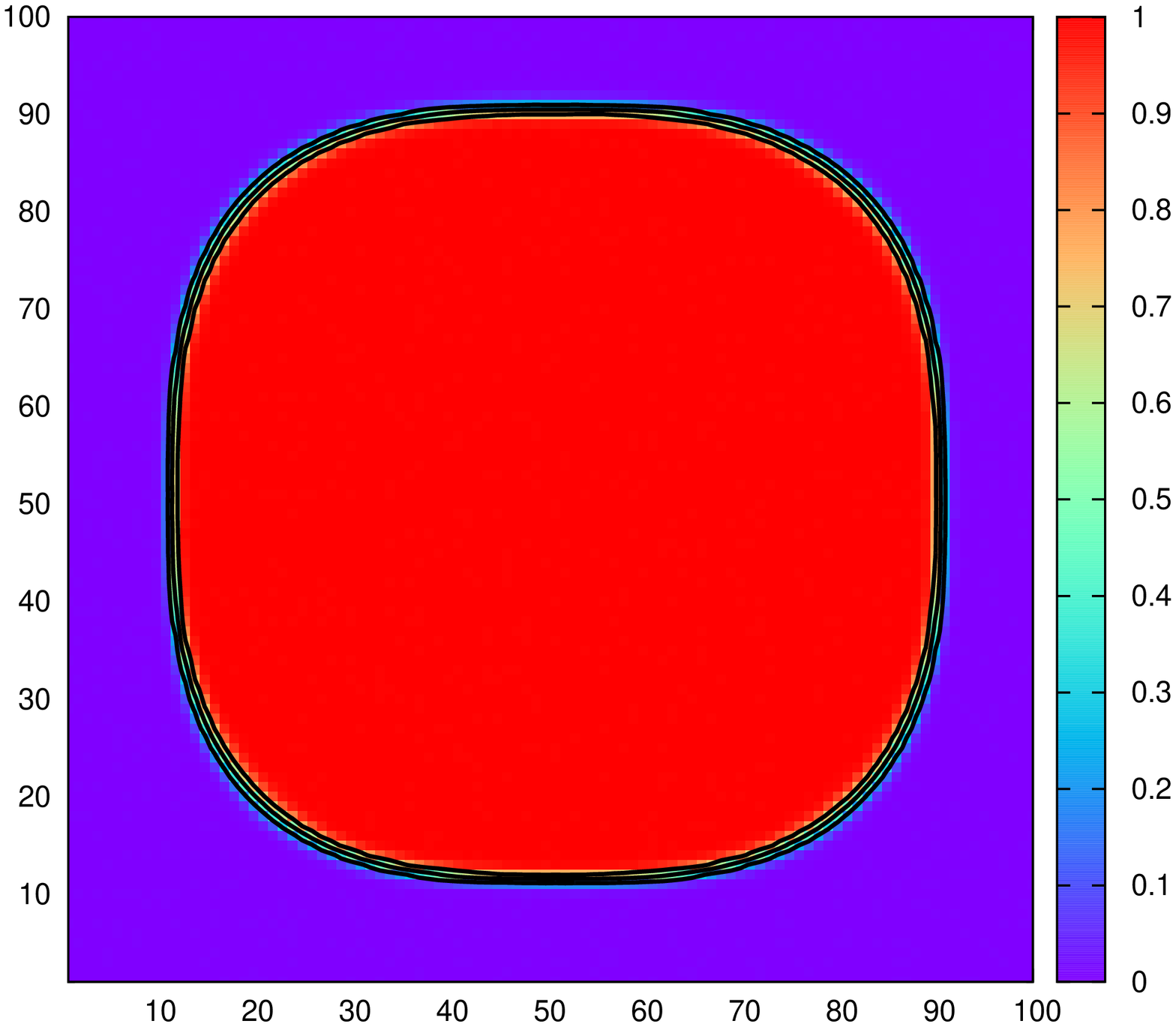}
	(c)
	\includegraphics[clip=true, trim=0cm 12mm 0mm 0cm, angle=-90,width=0.46 \textwidth]{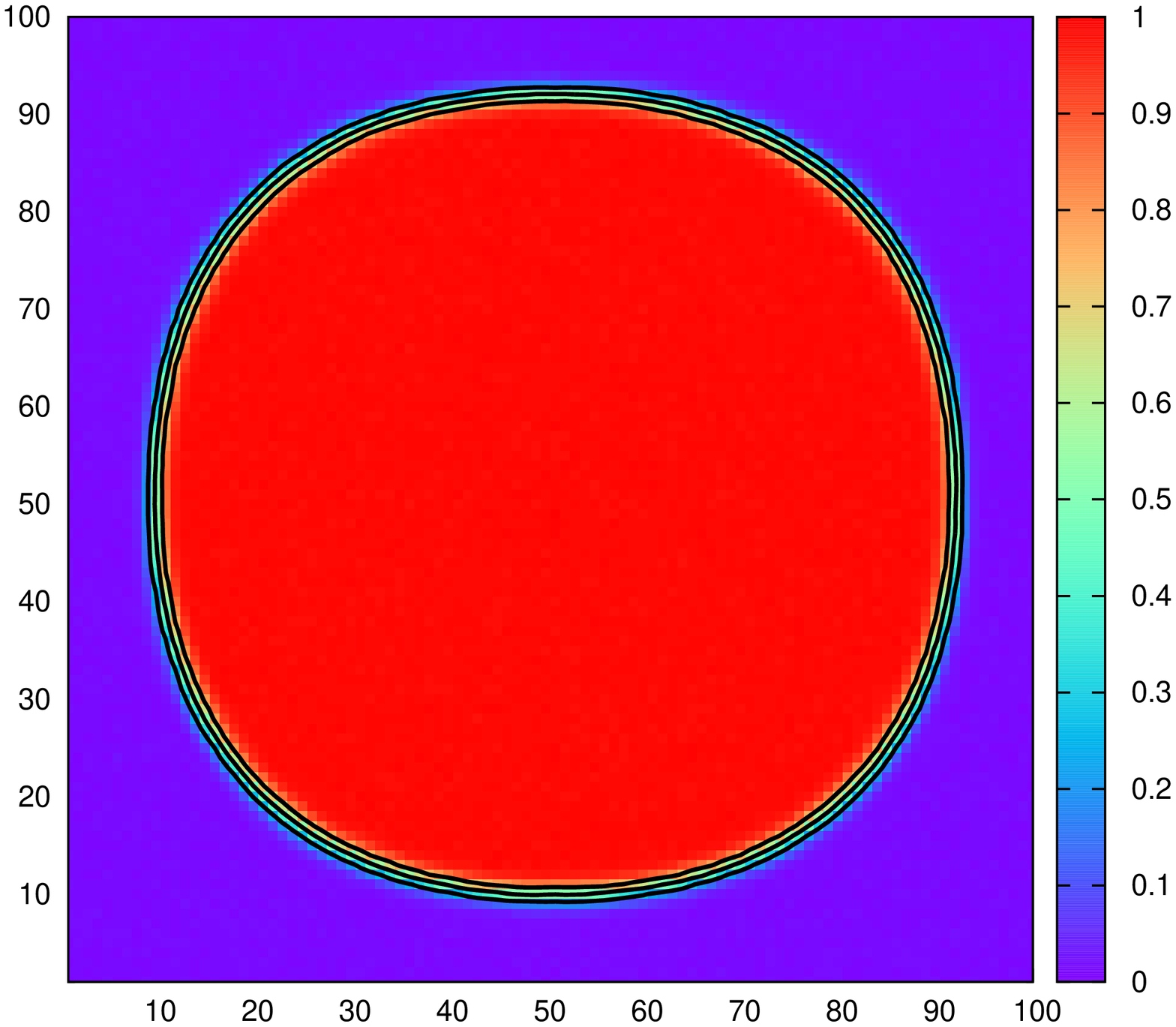}
	(d)
	\includegraphics[clip=true, trim=0cm 12mm 0mm 0cm, angle=-90,width=0.46 \textwidth]{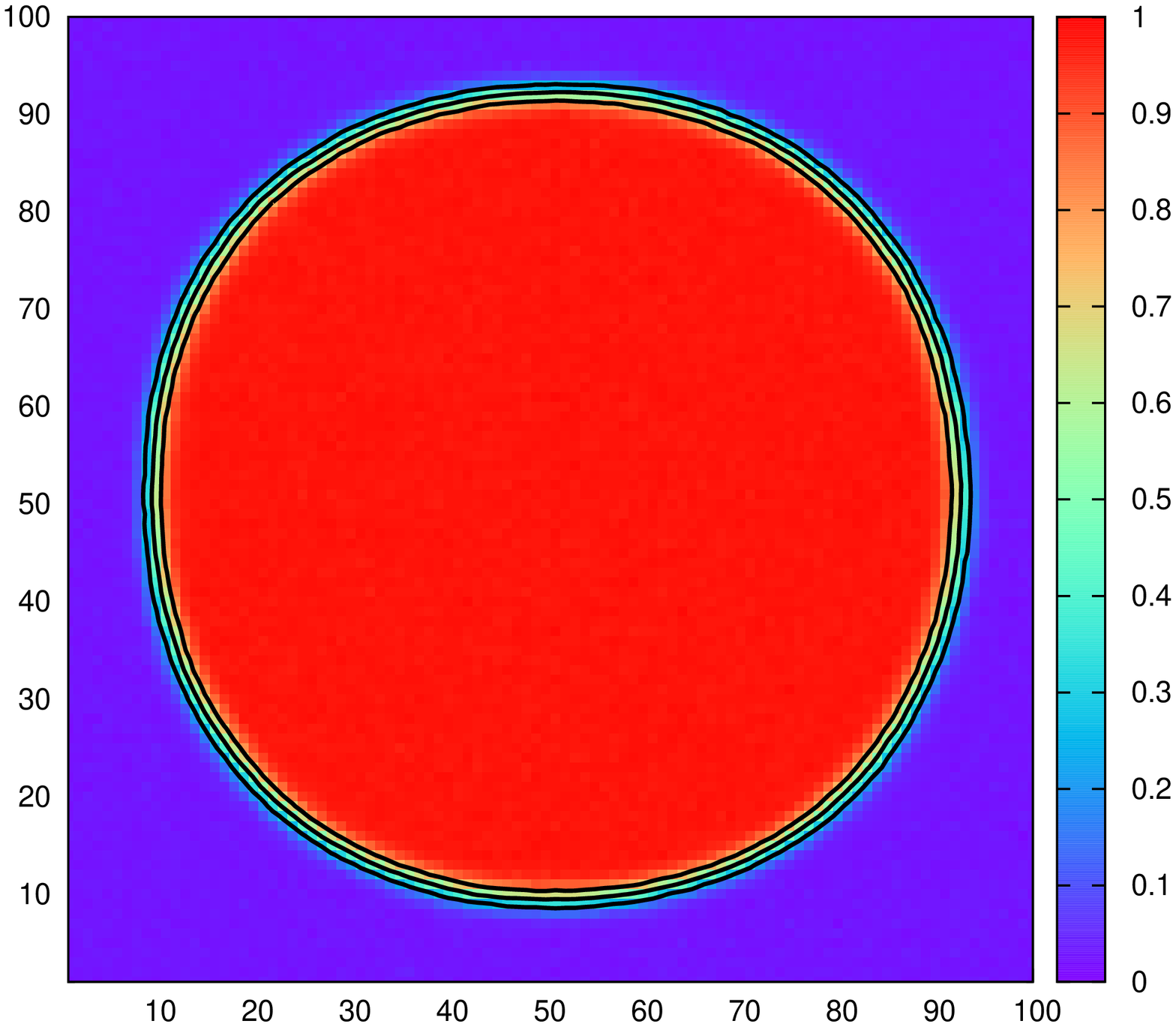}
	(e)
	\includegraphics[clip=true, trim=0cm 12mm 0mm 0cm, angle=-90,width=0.46 \textwidth]{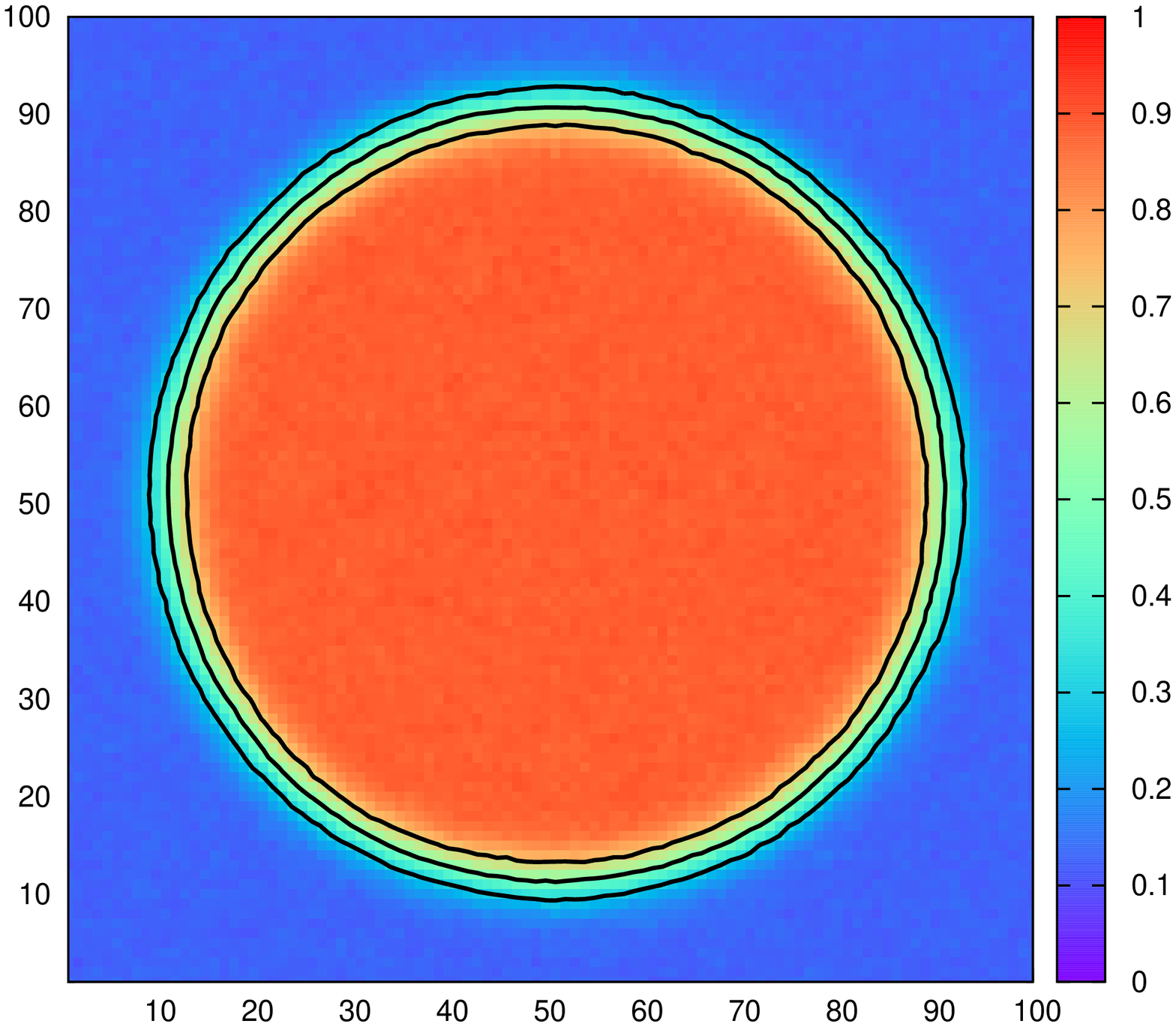}
	(f)
	\includegraphics[clip=true, trim=0cm 12mm 0mm 0cm, angle=-90,width=0.46 \textwidth]{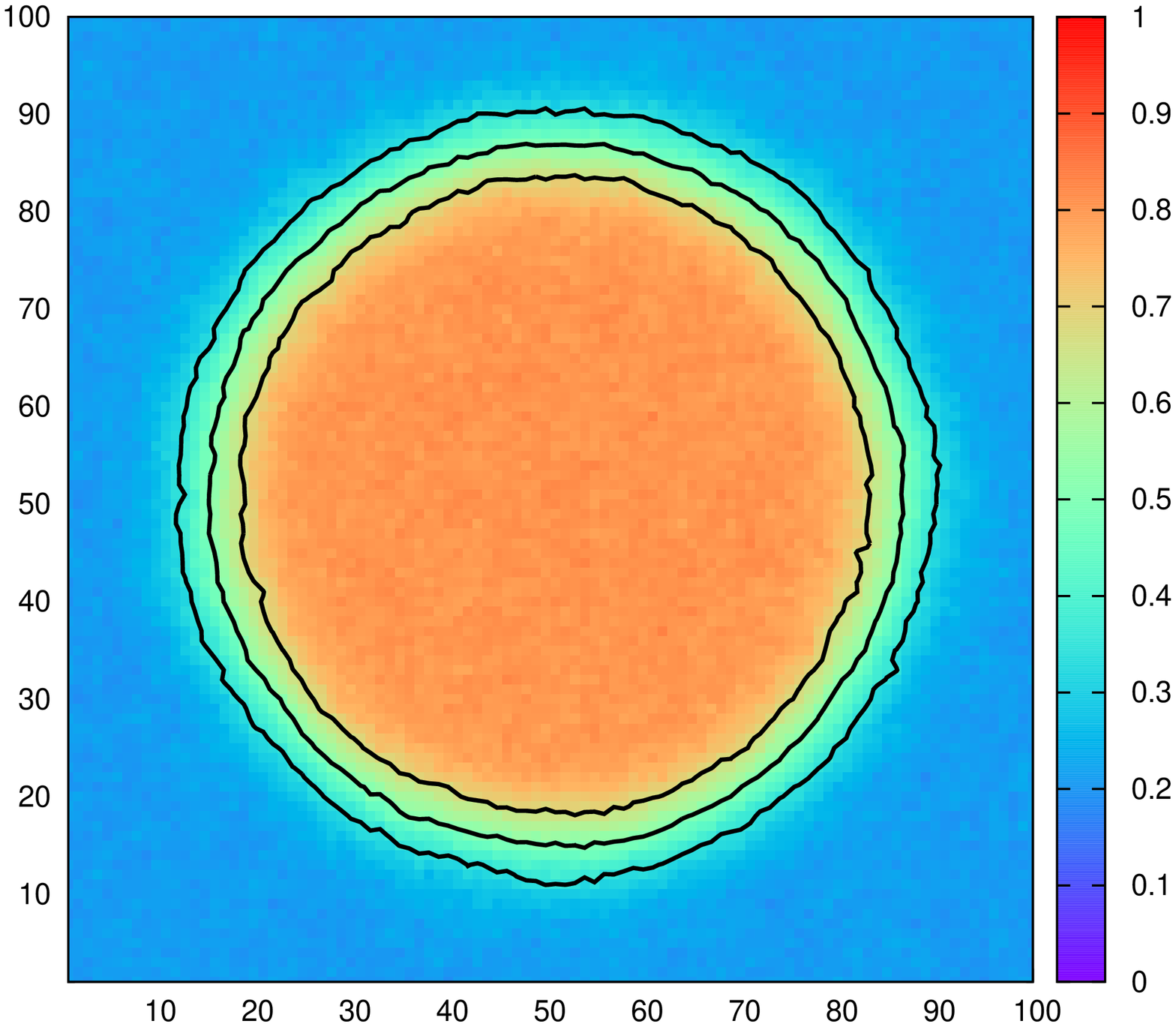}
	\caption{(Color online) Contours of constant density $\rho(x,y)=\rho_i$ in the planes $x=0$, $y=0$ and $z=0$ cutting through droplets situated at the origin in a box of dimension $L=100$ with periodic boundary conditions in every direction, for the cases $k_BT/J=1.0$ (a); $k_BT/J=2.0$ (b); $k_BT/J=2.6$ (c); $k_BT/J=3.0$ (d); $k_BT/J=4.0$ (e); $k_BT/J=4.3$ (f). The density is always fixed at $\rho=0.33$. In addition to the color-coded average density per site, the contours for three densities $\rho_i=(\rho_v(T)+0.5)/2, 0.5, (\rho_\ell(T)+0.5)/2$  are shown.}
\label{fig6}
	\end{center}
	\end{figure}

We deliberately do not discuss the radial droplet density profiles for $k_BT/J=2.0$ and $k_BT/J=2.6$, however, since at these temperatures the droplet shape shows distinct deviations from the spherical shape. This can be checked directly by recording contours of constant density in slices of width $= 1$ (taking into account 3 lattice planes through the droplet's center of mass, parallel to the $xy$ plane, the $xz$ plane and the $yz$ plane, respectively). Averaging these density contours over several hundred statistically independent observations the plots shown in Fig.~\ref{fig6} are obtained. They are all taken at the same fixed density $\rho=0.33$. Note that for this density, the stable state would be a cylindrical droplet (stabilized by the periodic boundary conditions) but for such large systems ($L=100$) the compact droplet shapes shown here (chosen by an appropriate initial condition) are always perfectly metastable. At $k_BT/J=2.0$ the cubic symmetry of these cross sections through the droplet is evident (although the presence of facets parallel to the planes $x=0$ or $y=0$ is not evident, due to finite-size effects at the points where the facets join the round sections, which replace the sharp edges of the cubes at nonzero temperature). At $k_BT/J=2.6$, where we exceed the roughening temperature slightly, there clearly occur no longer any facets, but the density contours in Fig.~\ref{fig6} are still distinctly non-circular: the diameter in diagonal direction clearly is about 7\% larger than in the lattice directions. Even at $k_BT/J=3.0$, we find an enhancement of the diameter in diagonal direction of about 3\%. At $k_BT/J=4.0$ to 4.3, however, no longer any statistically significant deviation from spherical droplet shapes (and hence circular shape of the density contours in the cross sections) can be detected. Note that for $k_BT/J=4.3$ (case (f)) the geometrical cluster definition would not be applicable due to the proximity of the percolation transition of geometrical clusters.

	\clearpage

	\begin{figure}[ht]
	\begin{center}
	(a)
	\includegraphics[clip=true, trim=0cm 12mm 0mm 0cm, angle=-90,width=0.46 \textwidth]{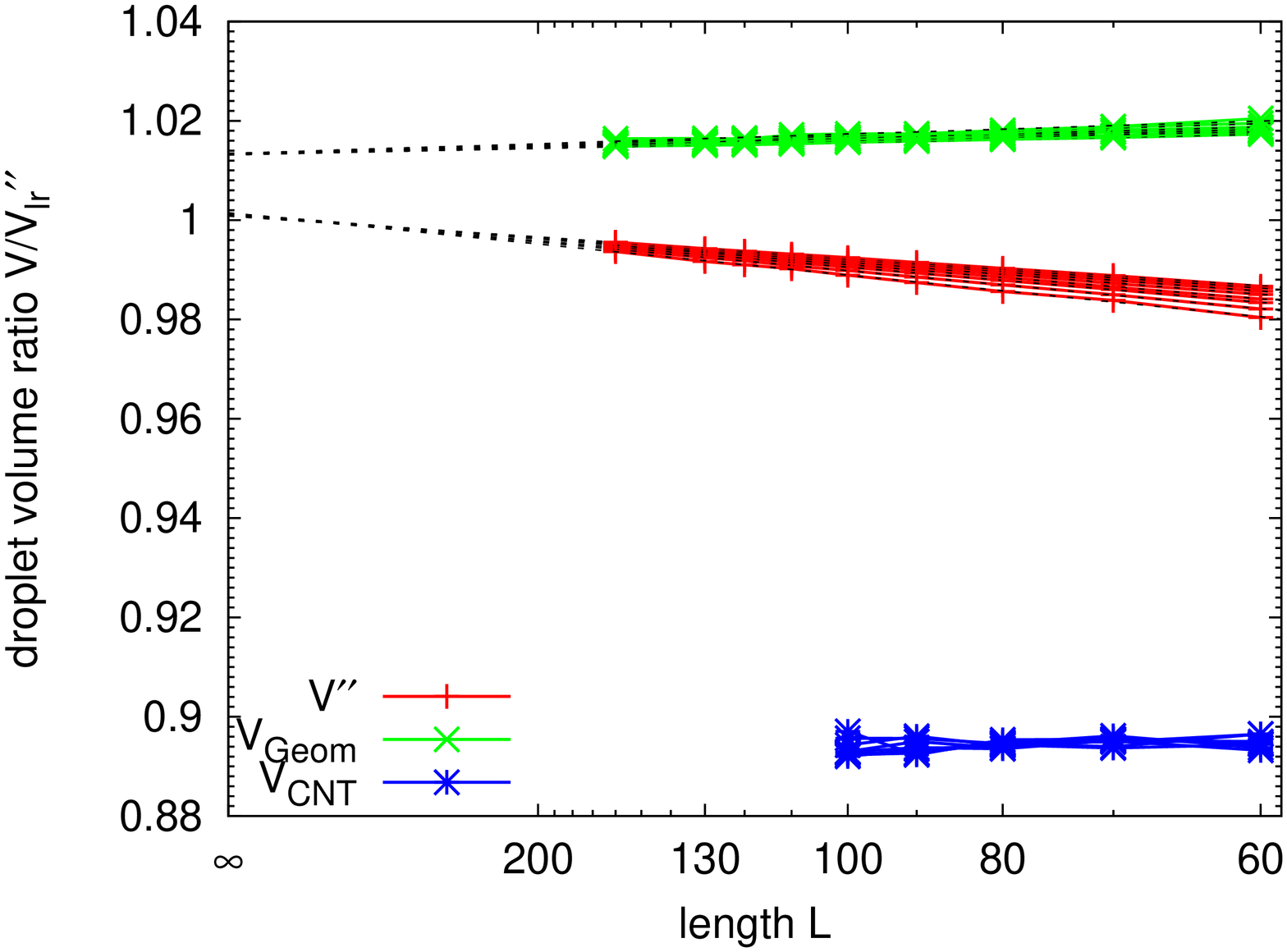}
	(b)
	\includegraphics[clip=true, trim=0cm 12mm 0mm 0cm, angle=-90,width=0.46 \textwidth]{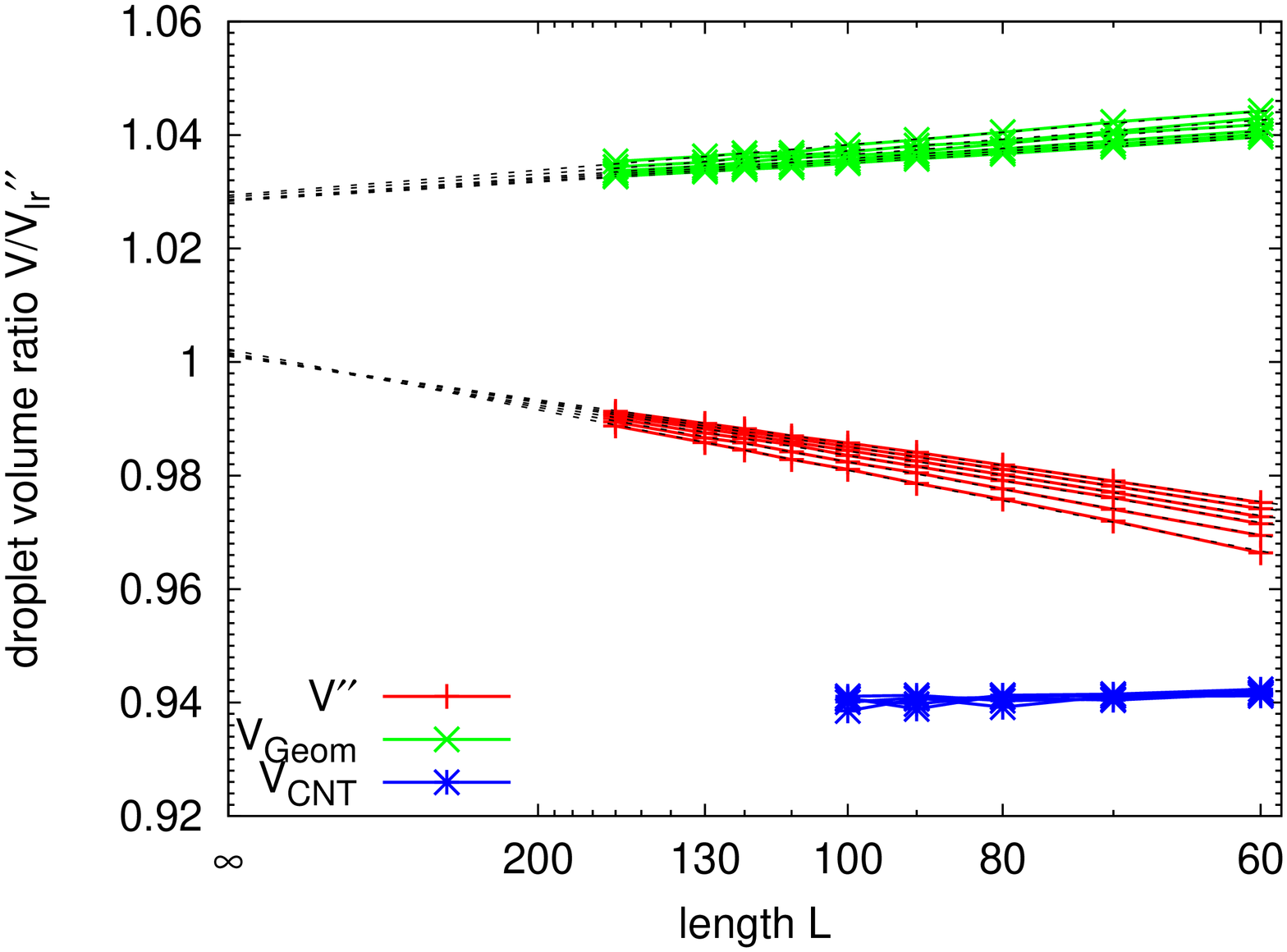}
	(c)
	\includegraphics[clip=true, trim=0cm 12mm 0mm 0cm, angle=-90,width=0.46 \textwidth]{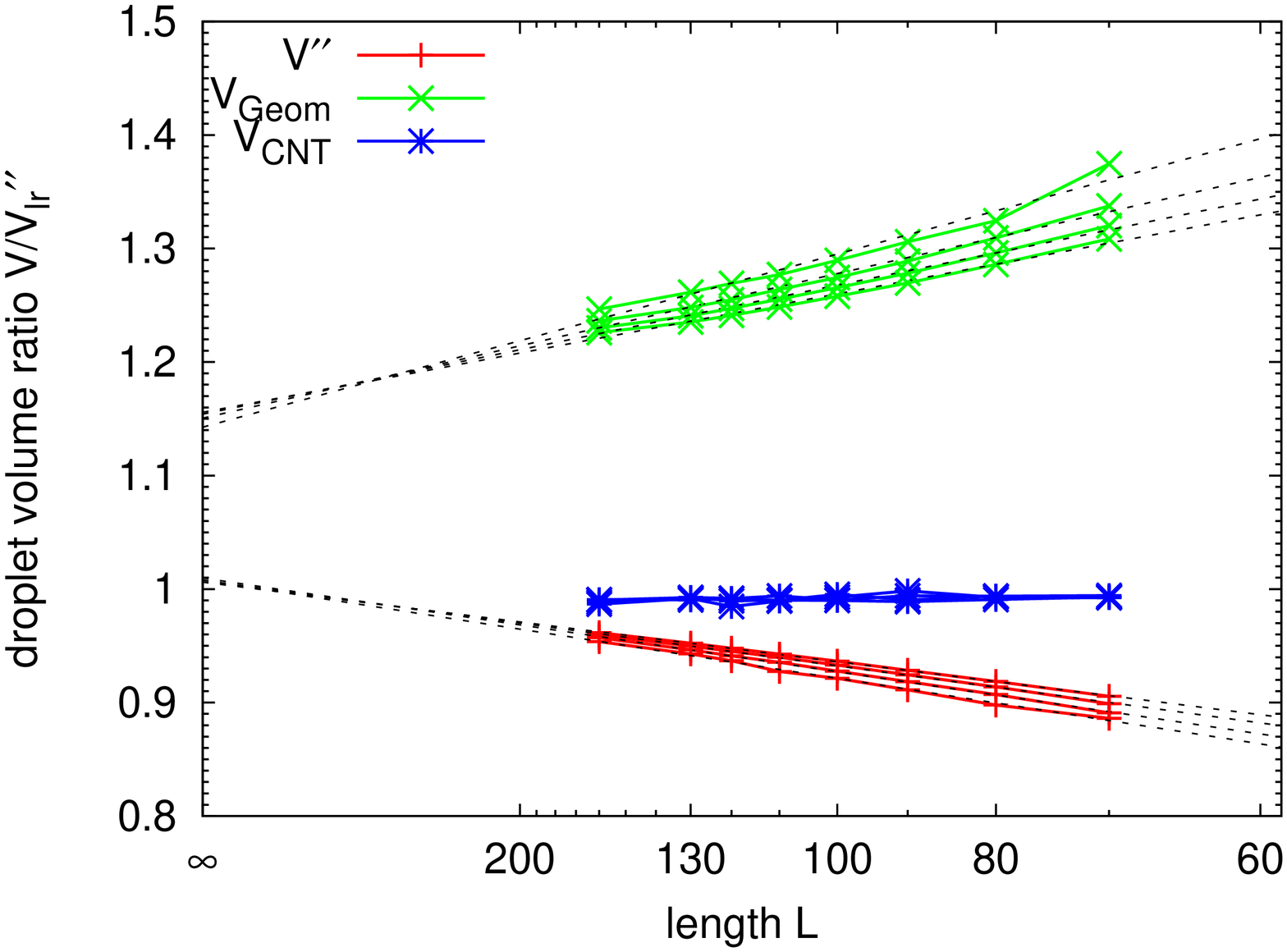}
	(d)
	\includegraphics[clip=true, trim=0cm 12mm 0mm 0cm, angle=-90,width=0.46 \textwidth]{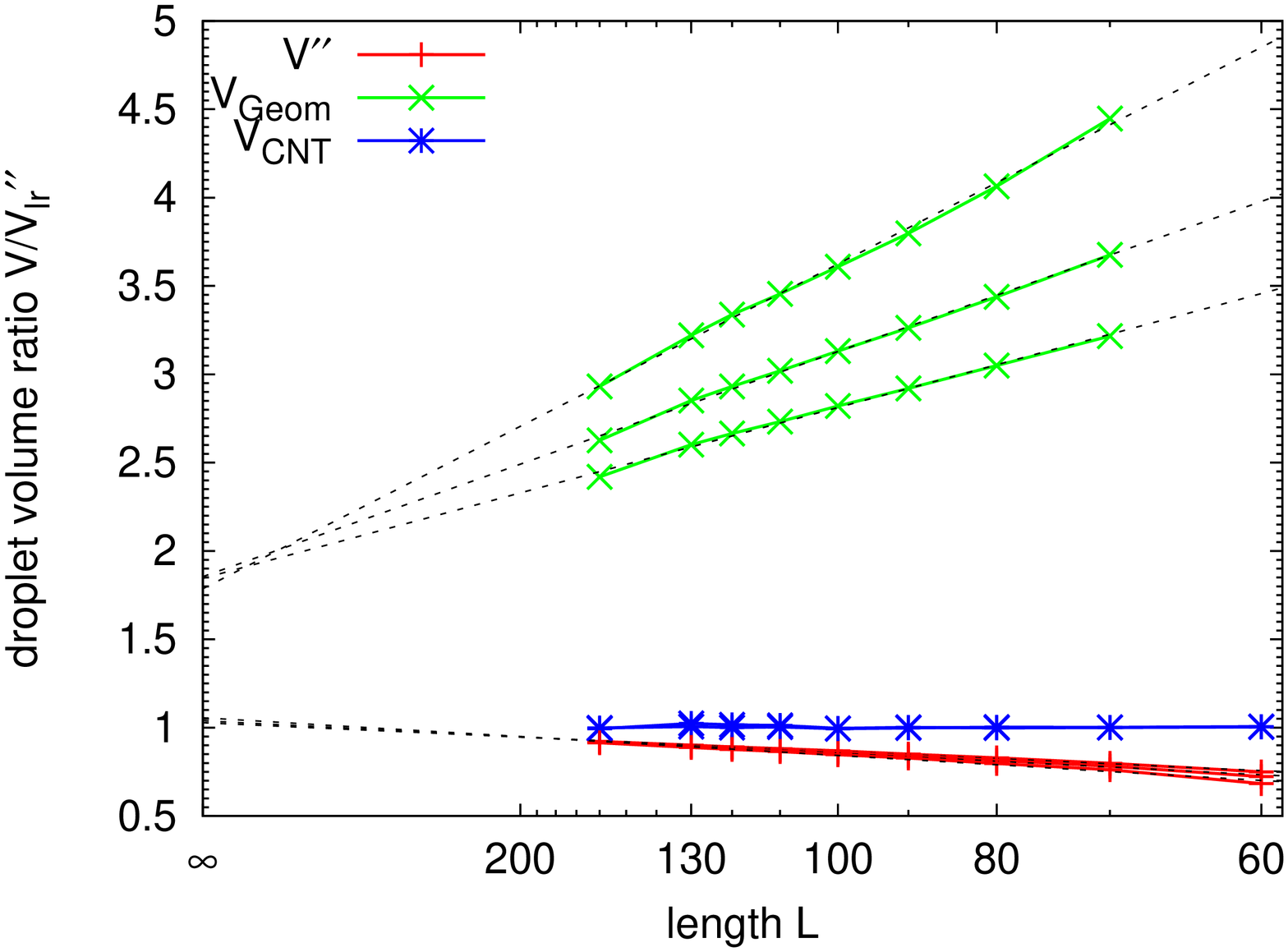}
	\caption{(Color online) Droplet volume $V''$ relative to the lever rule estimate $V_{lr}''$ plotted versus $1/L$ for four temperatures $k_BT/J=2.6$ (a), $3.0$ (b), $4.0$ (c) and $4.3$ (d). Three definitions of the droplet volume are compared to each other: the ``geometrical cluster'' definition $V_\text{geom}'' = 2 \langle \ell_\text{geom}\rangle/(1+m'')$, the ``physical cluster'' definition $V'' = \langle \ell\rangle/m''$ \{Eq. \eqref{eq9}\} and the result of classical nucleation theory $V'' = 4\pi (R^*)^3/3$ where $R^*$ is computed from \eqref{eq3}, using $\langle \Delta \mu \rangle$ as ``measured'' in the simulation by the lattice version of the Widom particle insertion method. At each temperature, several densities were used: $\rho=0.05, 0.06, \ldots, 0.12$ (a), $0.07, 0.08, \ldots, 0.12$ (b), $0.17, 0.18, \ldots, 0.20$ (c), $0.26, 0.27, 0.28$ (d).}
	\label{fig7}
	\end{center}
	\end{figure}

As a final part of our analysis of static properties of physical droplets in the Ising model, we exploit the fact that the chemical potential $\mu$ can be ``measured'' by the lattice version of the Widom particle insertion method \cite{55,84,85} also in the state when the system is inhomogeneous, e.g. for the case of interest when a droplet coexists with surrounding vapor. Actually, already in \cite{55} it was shown that $\mu$ actually stays spatially constant in such a situation. However, while the estimation of $\mu$ from Eq.~\eqref{eq6} requires a very accurate estimation of the free energy density $f_L(\rho,T)$ \{Eq.~\eqref{eq5}, Fig.~\ref{fig1}\}, and in practice this works only for not so large linear dimension $L$ (such as $L=20$ in Fig.~\ref{fig1}), the estimation of $\mu$ from the particle insertion method still works for volumes that are orders of magnitude larger. As a consequence, we can use Eq.~\eqref{eq3} to test whether or not the actual droplet volumes $V∞{''}$ are compatible with conventional nucleation theory for large droplets (assuming that the droplets are spherical) so $R^* = (3V''/4\pi)^{1/3}$ holds for a critical droplet.

Fig.~\ref{fig7} presents a plot of the cluster volume $V''$ versus inverse linear dimension for several densities, at the temperatures $k_BT/J=2.6$, $3.0$, $4.0$ and $4.3$, using the classical nucleation theory prediction $V_\text{CNT}''=4 \pi (R^*)^3/3$ with $R^*$ given by Eq.~\eqref{eq3} for comparison. For this purposes $\gamma_{v \ell}$ is taken from the work of Hasenbusch and Pinn \cite{79,100}, and so there occur no unknown parameters whatsoever. The geometrical cluster volume $V_\text{geom}''$ \{Eq. \eqref{eq: Vgeom}\} is close to the estimate based on the Coniglio-Klein-Swendsen-Wang ``physical cluster''-definition, Eq.~\eqref{eq9}, at $k_BT/J=2.6$ and $k_BT/J=3.0$, while at $k_BT/J=4.0$ (and higher) the deviations become appreciable: $V _\text{geom}''$ then is systematically too high (in comparison with all other estimates, including $V_{lr}''$ \{Eq.~\eqref{eq10}\}, which is used as a convenient normalization). It is interesting to observe that the classical nucleation theory estimates based on Eq.~\eqref{eq3} are systematically too low for $T=2.6$ and $T=3.0$, while for $T \geq 4.0$ Eqs.~\eqref{eq3}, \eqref{eq10} are found to be in very good agreement. This discrepancy at the relatively low temperatures comes from the fact that using Eq.~\eqref{eq3} for the estimation of $V''$ we imply that the cluster volume is spherical and hence we underestimate the surface area: the geometrical factor $C(T)$ introduced in Eq.~\eqref{eq11} increases from about 4.836 for the spherical shape near $T_c$ up to 6.0 for the cube, as the temperature is lowered. Since the non-spherical droplet shapes at $k_BT/J=2.6$ and 3.0 have regions of rather large curvature near the parts of the droplet where at low temperature the edges of the cube will appear, the estimation of $\mu$ then yields a too low radius $R^*$. The observation that the anisotropy of surface tension in the lattice gas model becomes noticeable for $k_BT/J < 4.0$ is consistent with the findings of Hasenbusch and Pinn \cite{79}.

If all the methods to define clusters were correct, in the thermodynamic limit all data should extrapolate to $V''/V_{lr}''=1$ for $L\to \infty$, since in this limit the lever rule \{Eq.~\eqref{eq10}\} is trivially true with $\rho'=\rho_v$ and $\rho''=\rho_\ell$. The method based on the definition of physical clusters, Eq.~\eqref{eq9}, indeed is nicely compatible with this expectation at all temperatures; although it is somewhat unsatisfactory (and unexpected) that at finite $L$ there occurs a surface correction due to the surface excess $N_\text{exc}$ noted in Eq.~\eqref{eq11} (and discussed above). However, it is clear that the two other methods do not give results that are correct for $L\to \infty$ in general: while the method based on the geometric cluster definition still gives essentially correct results at $k_BT/J=2.0$ (not shown here), where the distinction between ``geometrical'' and ``physical'' clusters is irrelevant, for temperatures $T>T_R$, the volume of geometrical clusters is systematically too large. At $k_BT/J=4.3$, the error is as large as 60 to 80\% even asymptotically, and for the cluster sizes that were actually studied the overestimation actually is by a factor two to five (Fig.~\ref{fig7}d)! In view of the fact that the geometric cluster definition must break down due to the percolation transition \cite{57}, this failure is not unexpected, but we are not aware that it ever has been quantified previously. In the regime from $k_BT/J=2.6$ to $4.0$, the error of the geometric cluster volume raises from a few percent to 15 to 40\%.

The results obtained from the classical nucleation theory via the ``measurement'' of the supersaturation $\Delta \mu$, on the other hand, yield essentially the correct result at high temperatures ($k_BT/J\geq4.0$), for all linear dimensions studied, since the data are essentially independent of $L$, and hence $R^*$, in the considered range. But it is remarkable that for $k_BT/J=3.0$ we find $V_\text{CNT}''/V_{lr}''\approx 0.94$ (Fig.~\ref{fig7}b) and for $k_BT/J=2.6$ we find $V_\text{CNT}''/V_{lr}''\approx 0.89$ (Fig.~\ref{fig7}a). This discrepancy becomes worse at lower temperatures (e.g.~$V_\text{CNT}''/V_{lr}''\approx 0.68$ at $k_BT/J=2.0$ [not shown]), and obviously this discrepancy must be attributed to the orientation dependence of the interfacial free energy and the resulting non-spherical droplet shapes (Fig.~\ref{fig6}).

\begin{figure}[ht]
	\begin{center}
	\includegraphics[clip=true, trim=0mm 5mm 0mm 0cm, angle=-90,width=0.68 \textwidth]{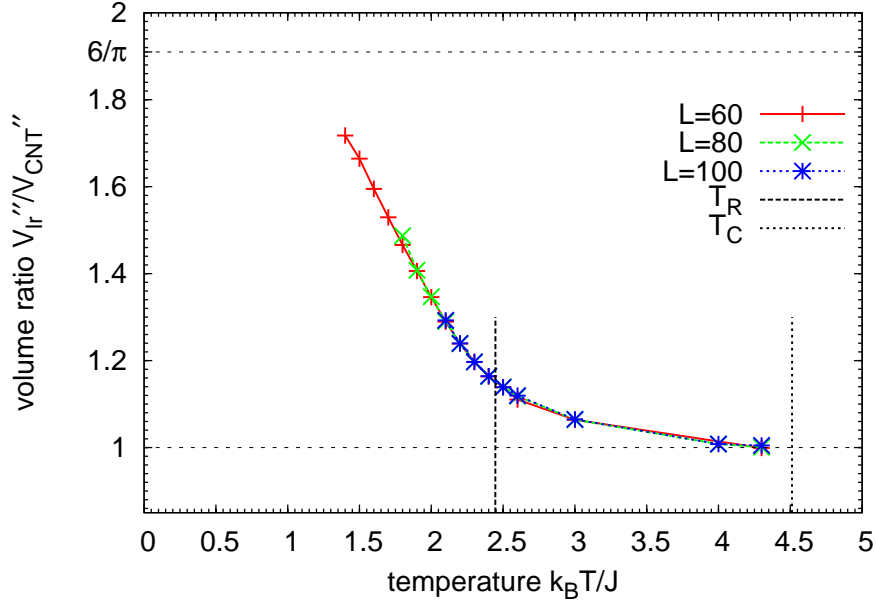}
	\caption{(Color online) Temperature variation of $V_{lr}''/V_\text{CNT}''$. The locations of the roughening temperature $T_R$ and the critical temperature $T_c$ are indicated by dotted vertical lines. As stated in the main text, the values of $\gamma_{vl}$ are taken from \cite{100}, using Monte Carlo results for temperatures above $k_BT/J=2.0$ and a low temperature series expansion up to 17th order for temperatures below $k_BT/J=2.0$. Note that for temperatures smaller than $k_BT/J=1.4$, the magnetization at coexistence $m_\text{coex}$ is almost indistinguishable from its saturation value $m_\text{coex}=1$, and then our implementation of the Widom particle insertion method cannot be applied to the chosen system sizes.}
	\label{fig: VCNTvsVLR}
	\end{center}
	\end{figure}

At first sight, the result that $V_\text{CNT}''/V_{lr}''$ is independent of $R^*$ seems to be at variance with the finding of a curvature-dependent surface tension $\gamma_{v\ell}(R)$ due to Winter et al. \cite{55,56,21,23,24,95}. In fact, evidence was provided that
	\begin{equation} \label{eq: GammaOfR}
	\gamma_{v\ell}(R) = \frac{\gamma_{v\ell}(\infty)}{1+2(l/R)^2}
	\end{equation}
where $l$ is a length proportional to the correlation length in the bulk \cite{95}. Note that due to the spin reversal symmetry of the Ising model one can show \cite{96} that a Tolman correction ($\propto 1/R$) must be absent for $R\to\infty$. However, from Eqs.~\eqref{eq2} and \eqref{eq: GammaOfR}, it is straightforward to show that
	\begin{equation}
	\begin{split}
	R^* \Delta \mu (\rho_\ell - \rho_v) &= 2 \gamma(\infty) \frac{1+4 (l/R^*)^2}{(1+2 (l/R^*)^2)^2} \\
	&=2 \gamma(\infty) \left[ 1-4(l/R^*)^4 + \mathcal{O}((l/R^*)^6) \right] \;.
	\end{split}
	\end{equation}
Hence for $R^*\gg l$ it is clear that the result for $R^*$ is still given by Eq.~\eqref{eq3}, and for $k_BT/J\leq4.0$ we are safely in this regime, since the correlation length then does not yet exceed the lattice spacing. So the deviations of the ratio $V_\text{CNT}''/V_{lr}''$ that are seen in Fig.~\ref{fig7} can be attributed fully to the deviation of the droplet shape from a perfect sphere, caused by the anisotropy of the interfacial free energy. In Fig.~\ref{fig: VCNTvsVLR}, we now present the ratio of the intercepts $V_\text{lr}''/V_{CNT}''$ for $R^*\to\infty$ as a function of temperature, since we know that \{Eq.~\eqref{eq3}\} $V_\text{CNT}^* = (4\pi/3) \gamma_{v \ell}^3(\infty)/[\Delta \mu(\rho_\ell-\rho_v)]^3$, while Eq.~\eqref{eq4a} yielded $\Delta F_\text{hom}^*/k_BT = (36\pi)^{1/3}  \gamma_{v \ell}(\infty)/3 (V_\text{CNT}^*)^{2/3} = m_\text{coex}H V_\text{CNT}^*$ for a spherical droplet. At $T=0$, however, the droplet is a perfect cube, and for intermediate temperatures, its shape (for $V^*\to \infty$) is given by the Wulff construction \cite{101} (and hence not explicitly known). However, for large droplet volume $V$ we can write in general
	\begin{equation}
	\Delta F(V) = -2m_\text{coex} H V + \gamma \widetilde{A}\widetilde{V}^{-2/3} V^{2/3} \;,\qquad V\to \infty
	\end{equation}
when we have assumed that the droplets of different linear dimension $R$ (for $R\to \infty$) have the same shape at fixed temperature, so we can write $V=\widetilde{V}R^3$ for the droplet volume ($\widetilde{V}$ is then formally the volume for $R=1$) and the surface area is $A=\widetilde{A}R^2$. For instance, for a sphere we have $\widetilde{V}_\text{sphere} = 4\pi/3$ and $\widetilde{A}_\text{sphere} = 4\pi$, and for the cube $\widetilde{V}_\text{cube}=1$ and $\widetilde{A}=6$. Minimizing $\Delta F(V)$ with respect to $V$ yields
	\begin{equation}
	(V^*)^{1/3} = \frac13 \frac{\gamma}{m_\text{coex}H} \widetilde{A}\widetilde{V}^{-2/3}
	\end{equation}
for a general shape, which is in between sphere and cube. The barrier then can be written as
	\begin{equation} \label{eq: DeltaFGeneral}
	\Delta F^* = \frac{1}{27} \frac{\gamma^3}{(m_\text{coex}H)^2} \widetilde{A}^3 \widetilde{V}^{-2} = m_\text{coex} H V^*\; .
	\end{equation}
Using now the fact that by choosing a particular large droplet volume in our simulation, $H$ is automatically fixed for any volume $V_{lr}^*$ due to the thermal equilibrium situation constructed in our simulation. So it makes sense to estimate the ratio of barriers as
	\begin{equation}
	\frac{\Delta F^*}{\Delta F_\text{CNT}^*} = \frac{V_{lr}^*}{V_\text{CNT}^*}
	\end{equation}
For $T=0$, we know that $\gamma$ is again the interface tension of the planar surface, also used in the classical nucleation theory for the spherical surface. Hence when we write the ratio of $\Delta F^* / \Delta F_\text{CNT}^*$, using Eq.~\eqref{eq: DeltaFGeneral}, the term $\gamma^3/(27(m_\text{coex}H)^2)$ cancels,
	\begin{equation}
	\frac{\Delta F_{T=0}^*}{\Delta F_\text{CNT}^*} = \frac{\widetilde{A}_\text{cube}^3 \widetilde{V}_\text{cube}^{-2}}{\widetilde{A}_\text{sphere}^3 \widetilde{V}_\text{sphere}^{-2}} = \frac{6}{\pi}\; .
	\end{equation}
This asymptotic value of $\frac{V_{lr}^*}{V_\text{CNT}^*}$ should be reached for $T\to 0$, while $\frac{V_{lr}^*}{V_\text{CNT}^*}\to 1$ as $T\to T_c$. Our numerical results (Fig.~\ref{fig: VCNTvsVLR}) are compatible with this expectation.

\clearpage
 \section{Time evolution of the droplet size distribution and droplet growth rates}
 \label{sec3}

 We now consider time-dependence of the system where we start the system at time $t=0$ in an equilibrated state in the vapor at $\mu = \mu_\text{coex}$, but switch on at time $t=0$ a chemical potential $\mu>\mu_\text{coex}$ (or equivalently, a positive magnetic field $H$ in the notation of Ising ferromagnet) at which the liquid is the stable phase. As mentioned in the introduction, we consider heterogeneous in addition to homogeneous nucleation, choosing a $L \times L \times D$ geometry with two walls at $z=1$ and $z=D$, choosing surface fields $H_1,\; H_D$ such that $H_D$ favors the vapor but $H_1$ (acting at the wall at $z=1$) favors the liquid. The reason for this choice is, that by proper choice of $H_1$ one can adjust the contact angle $\theta$ at which sessile wall-attached macroscopic droplets can occur. The barrier against heterogeneous nucleation $\Delta F_\text{het}^*$ is predicted to be very much reduced, in comparison to the barrier $\Delta F_\text{hom}^*$ against homogeneous nucleation, if the contact angle is small, since \cite{28,29}
\begin{align}\label{eq14}
\begin{split}
\Delta F^*_\text{het} &= \Delta F_\text{hom}^* f(\theta)\;,\\
 f(\theta) &= (1-\cos \theta)^2(2+\cos \theta)/4\; .
 \end{split}
\end{align}
As shown with the ``lever rule'' method \cite{55,56} indeed rather large wall attached droplets can be simulated in equilibrium with supersaturated vapor which have barriers of order $\Delta F_\text{het}^* \approx 10 k_BT$ or so only, for suitable choices of $H_1$ and $H$, and so a comparison with kinetic studies then seems reachable, and varying $H_1$ over some range provides an additional variable to test the theory.

	\begin{figure}[ht]
	\begin{center}
	(a)
	\includegraphics[clip=true, trim=0cm 12mm 8mm 0cm, angle=-90,width=0.46 \textwidth]{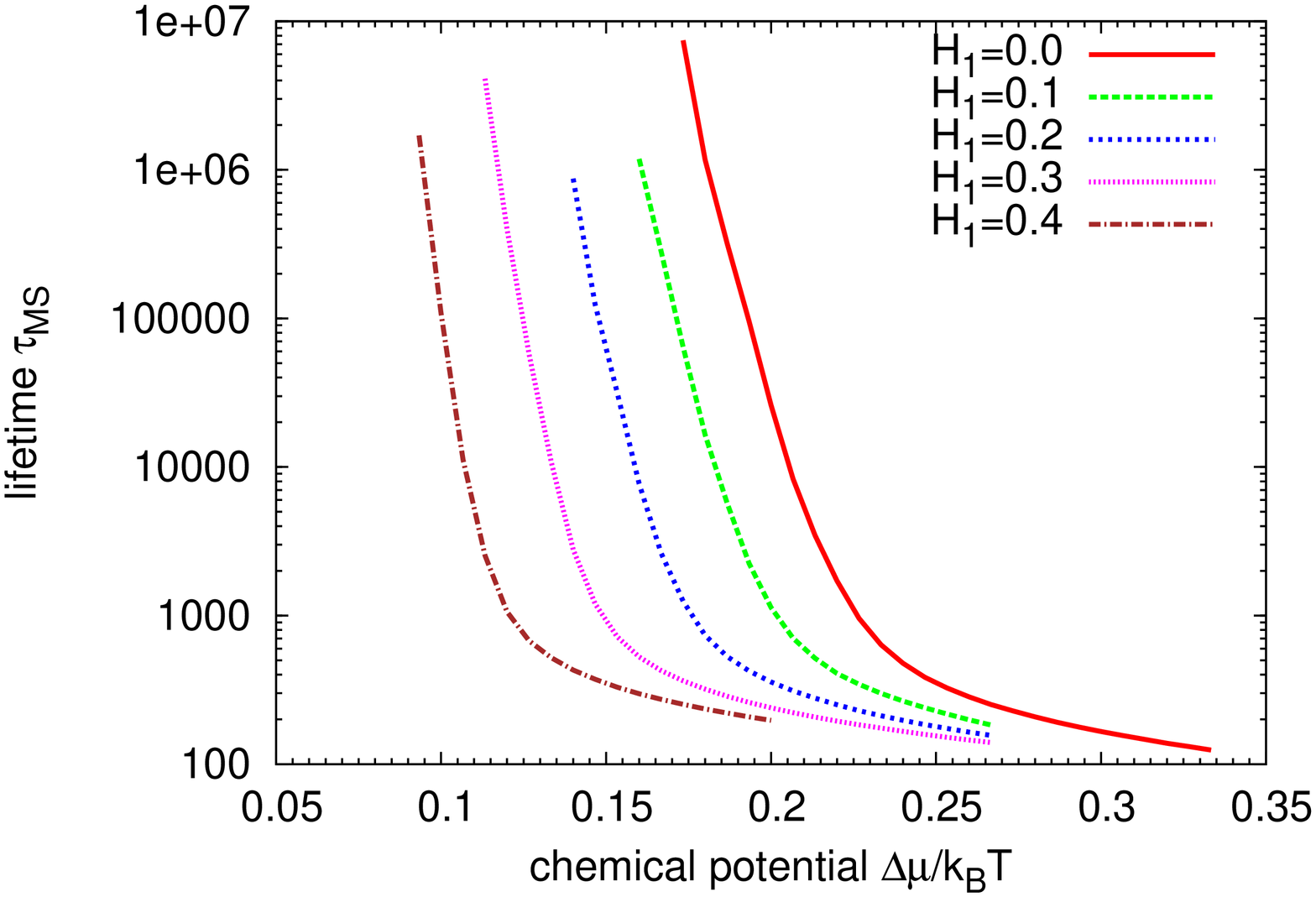}
	(b)
	\includegraphics[clip=true, trim=0cm 12mm 8mm 0cm, angle=-90,width=0.46 \textwidth]{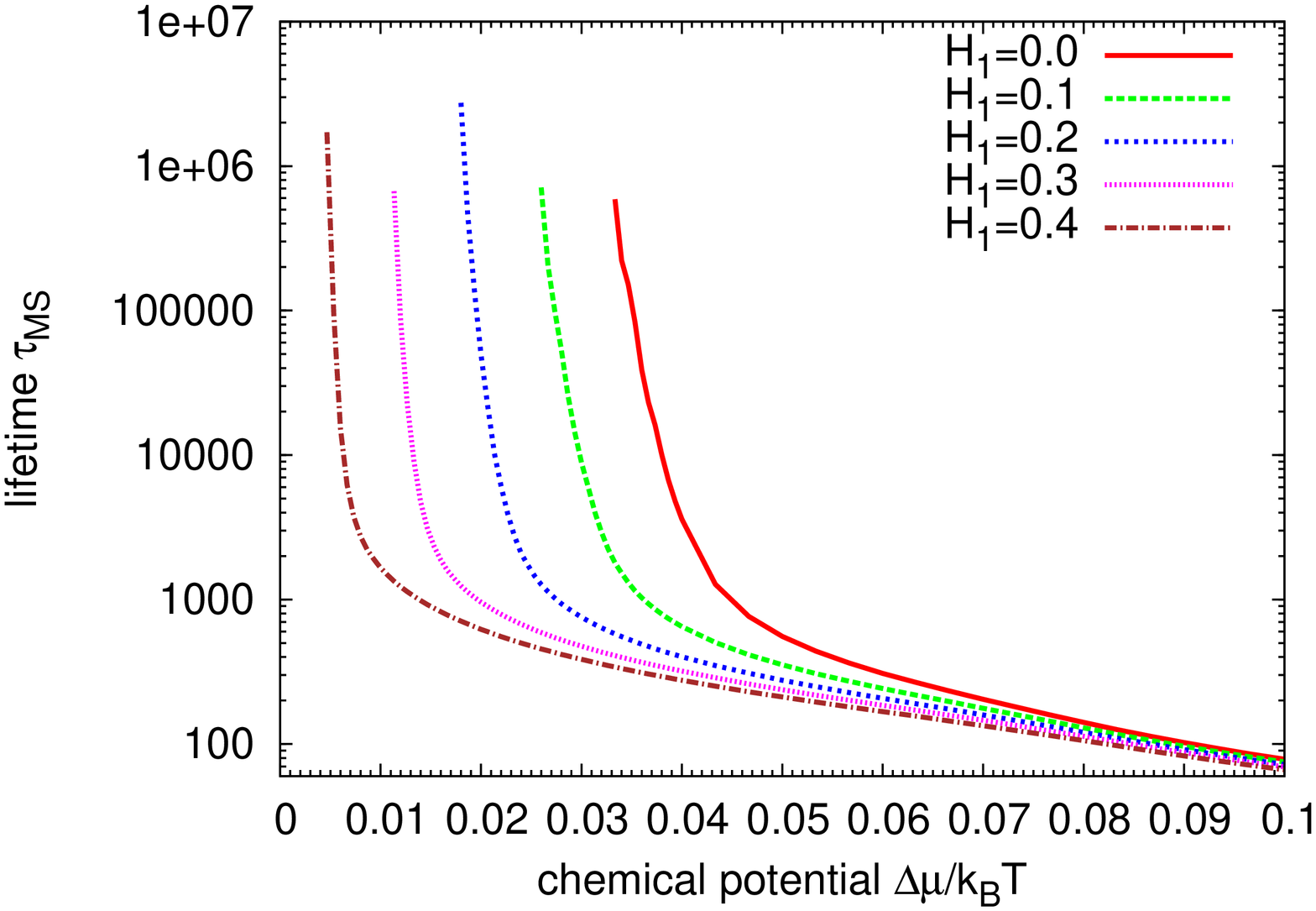}
	\caption{(Color online) Mean lifetime of metastable states in $L\times L \times D$ Ising systems at $k_BT/J=3.0$ (a) and $k_BT/J=4.0$ (b) as a function of the chemical potential $\mu/k_BT$, for various choices of $H_1$ as indicated. This lifetime was measured as the first time, when the time-dependent magnetization $m(t)$ becomes positive, as described in more detail in the main text. $L=60$ and $D=30$, with $H_D/J=-0.9$ throughout, and periodic boundary conditions in $x$ and $y$ directions only. The average lifetime decreases with higher  surface fields $H_1$ and highter chemical potentials $\Delta \mu/k_BT$.}
	\label{fig8}
	\end{center}
	\end{figure}

As a first step, preliminary runs were performed with the single spin flip Metropolis algorithm \cite{80,81} monitoring the average lifetime of the metastable vapor. This was done using a large sample $(10^5$) of equilibrated initial states at $H=0$, where the considered field $H$ (or chemical potential $\mu - \mu_\text{coex}$, respectively) was then switched on and the time recorded when the (initially negative) magnetization reaches the value $m=0$ for the first time. Fig.~\ref{fig8} shows estimates for the resulting mean first passage times for a range of choices of $H_1$ as a function of the field. When this ``lifetime'' of the state with $m <0$ (i.e., vapor) does not exceed $10^4$, the system is rather unstable, nucleation occurs fast and is followed by fast domain growth as well; such fast decays of unstable systems are not suitable for tests of nucleation theory. It is seen, that in a rather narrow interval of fields $H$ (for each value of $H_1$) the lifetime increases from $10^4$ to $10^6$. Such parameter combinations $(H,H_1)$ will be studied in the following only; if we would study cases where the lifetime is significantly larger than $10^6$, no critical droplet would be formed during affordable simulation times.

	\begin{figure}[ht]
	\begin{center}
	(a)
	\includegraphics[clip=true, trim=0cm 12mm 0mm 0cm, angle=-90,width=0.46 \textwidth]{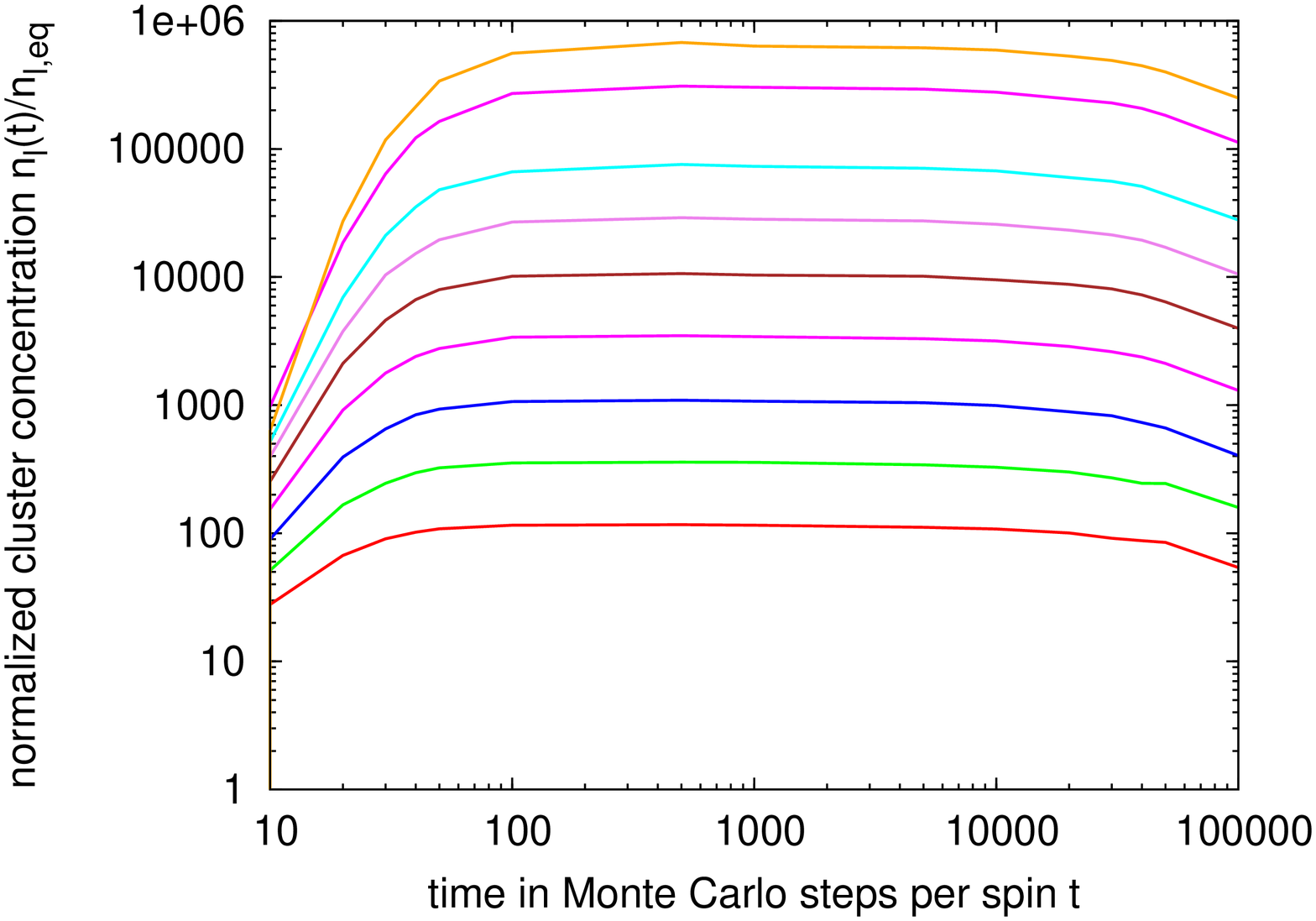}
	(b)
	\includegraphics[clip=true, trim=0cm 12mm 0mm 0cm, angle=-90,width=0.46 \textwidth]{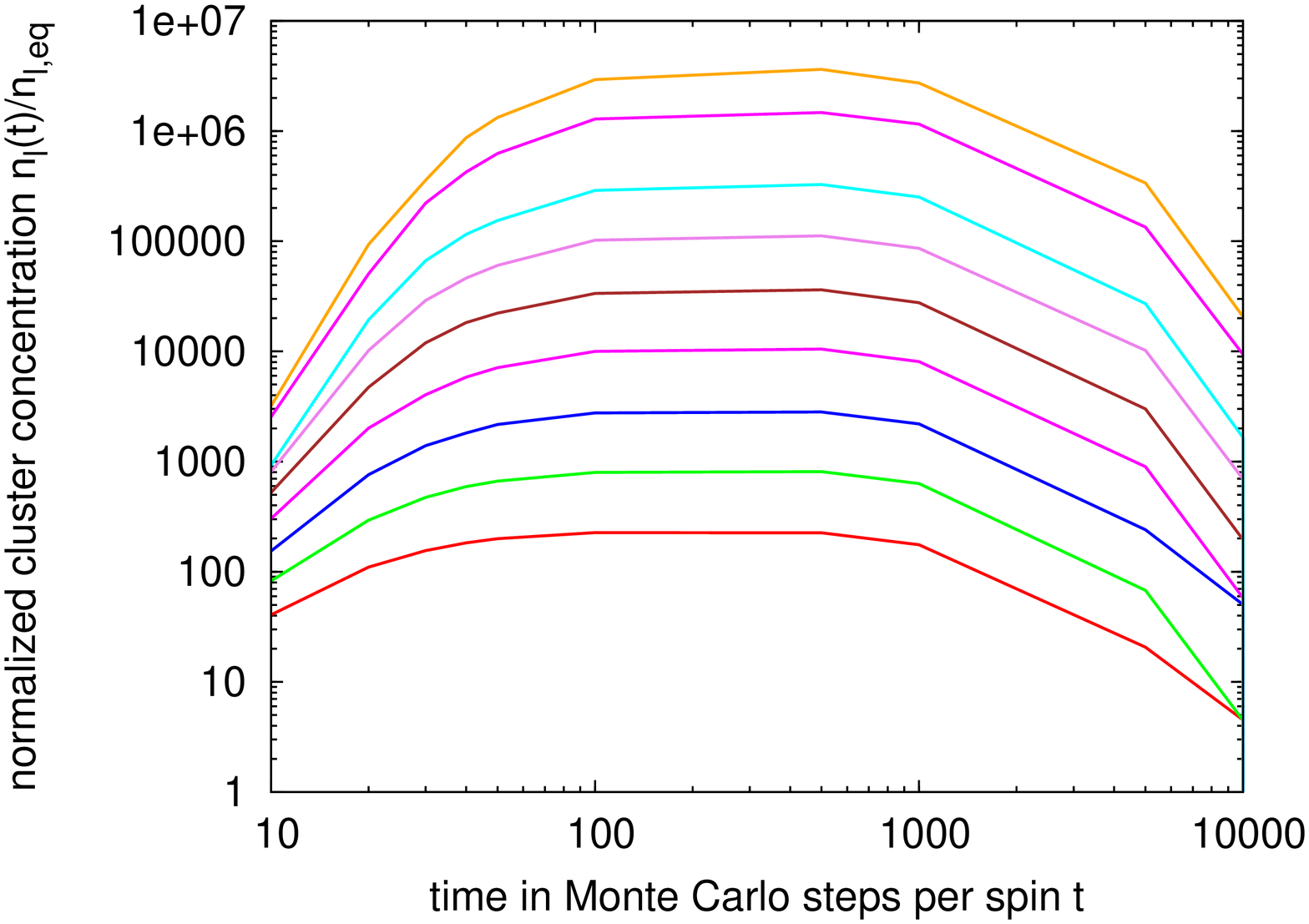}
	\caption{(Color online) Time dependence of the cluster concentration ratio $n_\ell(t)/n_\ell^{eq}$ in equilibrium at $H=0$, for the case $k_BT/J=3$, $H_1/J=0.4$, and $H/J=0.15$ (a) and $0.17$ (b). Curves show the choices $\ell = 40,50, \ldots, 120$ (from bottom to top).}
	\label{fig9}
	\end{center}
	\end{figure}

Fig.~\ref{fig9} then shows typical time evolutions of the size distribution of $n_\ell(t)$ of physical clusters of size $\ell$, normalizing them by the equilibrium cluster concentration $n_\ell^\text{eq}$ for $H=0$. Here $n_\ell^\text{eq}$ is defined as the average number of physical clusters per lattice site, and we have the sum rule
\begin{equation}\label{eq15}
\rho= \sum \limits _{\ell =1}^N \ell n _\ell^\text{eq}
\end{equation}
since every site occupied by a particle must be part of some cluster. Of course, here we are only interested in not too small clusters, and hence Fig.~\ref{fig9} focuses on clusters with $\ell \geq 40$. In the time evolution of $n_\ell(t)/n_\ell^\text{eq}$ we recognize three regimes: for times of order $t \leq 100, \; n_\ell(t)/n_\ell^\text{eq}$ is rapidly rising: this period of time corresponds to the relaxation from the initial state (where $H=0$) towards the metastable state. In the latter, $n_\ell(t)$ is almost constant for at least one, or even several decades of time. Then a decay of these plateau values sets in, which is due to the fact that too many much larger clusters have grown, the volume fraction of the system that is still in the metastable phase shrinks, and so less clusters of intermediate size (as studied in Fig.~\ref{fig9}) are observed. This behavior is qualitatively similar to previous studies (e.g. \cite{9,33}) which were based on the geometrical cluster definition, however. We also remark that in the case of Fig.~\ref{fig9}b the lifetime of the plateau extends up to $t \approx 1000$, one decade only, as expected from Fig.~\ref{fig8}a, since this case corresponds to a ``lifetime'' of the metastable state of only $\tau_{MS} \approx 3000$, and it is clear that only times $t$ distinctly less than $\tau _{MS}$ should be analyzed.

While some of the previous work on the studies of the kinetics of cluster growth in metastable Ising models (e.g. \cite{9,33}) tried to use directly $n_\ell(t)$ to extract information on the validity of nucleation theory concepts, we here try to implement a different concept. Namely, we follow the trajectories of individual (large) clusters with respect to their size in time, $\{\ell_i(t)\} \rightarrow \{\ell _i '(t + \Delta t)\} \rightarrow \ldots$, where $i$ is an index to label individual clusters. To ensure that the $i$'th cluster at time $t+\Delta t$ is actually a descendent of the $i$'th cluster at the time $t$, we have to choose $\Delta t$ small enough, and also record the location (center of gravity $\vec{X}_i(t))$ and the components of the gyration radius
\begin{equation}\label{eq16}
R_{i, \alpha}(t)=\left\{ \frac{1}{\ell_i(t)-1} \left(\sum_{k=1}^{\ell_i(t)} x_{k,\alpha}^2 - \frac{1}{\ell_i(t)} (\sum_{k=1}^{\ell_i(t)}x_{k,\alpha})^2\right)\right\}^{1/2}\;,
\end{equation}
where $x_{k,\alpha}$ is the $\alpha$'th Cartesian coordinate for the $k$'th lattice site belonging to the cluster with label $i$, consisting of $\ell_i(t)$ lattice sites. In each time step in which an analysis of the clusters is performed, the set of coordinates $\{\vec{X}_i(t),\vec{R}_i(t)\}$ is recorded.

Note that there occurs the difficulty that the number of large clusters is not constant during the simulation: clusters form and decay or split into parts, and since we know that $\ell_\text{geom}$ exceeds $\ell$ for each cluster, and the assignment of the ``active bonds'' according to Eq.~\eqref{eq4} to identify from the geometrical cluster the associate physical clusters is a random process, some random shift of $\vec{X}_i(t)$ would occur even if we carry out two successive cluster identifications from the same spin configuration $(\Delta t=0$). Of course, such shifts should be small in comparison with $\vec{R}_i(t)$, but as $\Delta t$ is chosen nonzero it is clear that useful results are only obtained if $\Delta t$ is small enough, and $\ell$ is large in comparison to clusters that correspond to typical thermal fluctuations \cite{58,61}. Hence only clusters for which $\ell >\ell_\text{min}$ are considered (for the temperature $k_BT/J=3.0$ we chose arbitrarily $\ell_\text{min}=10$). So if by such criteria (for details see \cite{86}) it is ensured that the $i$'th cluster with size $\ell_i '(t+\Delta t)$ is a descendent of the $i$'th cluster with size $\ell_i(t)$ at time $t$, we can define a reaction rate $\Gamma (\ell) $ as
\begin{equation}\label{eq17}
\Gamma (\ell) = \left\langle \frac{\ell_i'(t+\Delta t) - \ell_i(t)]}{\Delta t} \right\rangle _{\ell_i(t)}
\end{equation}
Here the index $\ell_i(t)$ stands for an average over a sampling of all cluster trajectories recorded in the simulation, and a smoothing procedure of the (otherwise too noisy) data with a triangular smoothing function \cite{86} was applied. Of course, in order to collect statistically significant data on $\Gamma (\ell)$, it is necessary to perform many runs for each parameter combination $(T,H,H_1)$ that is studied. We observe that the lifetime of the metastable stale in such runs is fluctuating dramatically, and so it is necessary to choose the run time of each run individually, rather than the same for all runs. It was decided to stop each run automatically when the largest cluster size $\ell^\text{max}_i(t) = L^3/20$ was reached. Of course, then only clusters with $\ell \ll L^3/20$ could be studied, to avoid artifacts caused by this cutoff.

	\begin{figure}[ht]
	\begin{center}
	(a)
	\includegraphics[clip=true, trim=0cm 12mm 0mm 0cm, angle=-90,width=0.46 \textwidth]{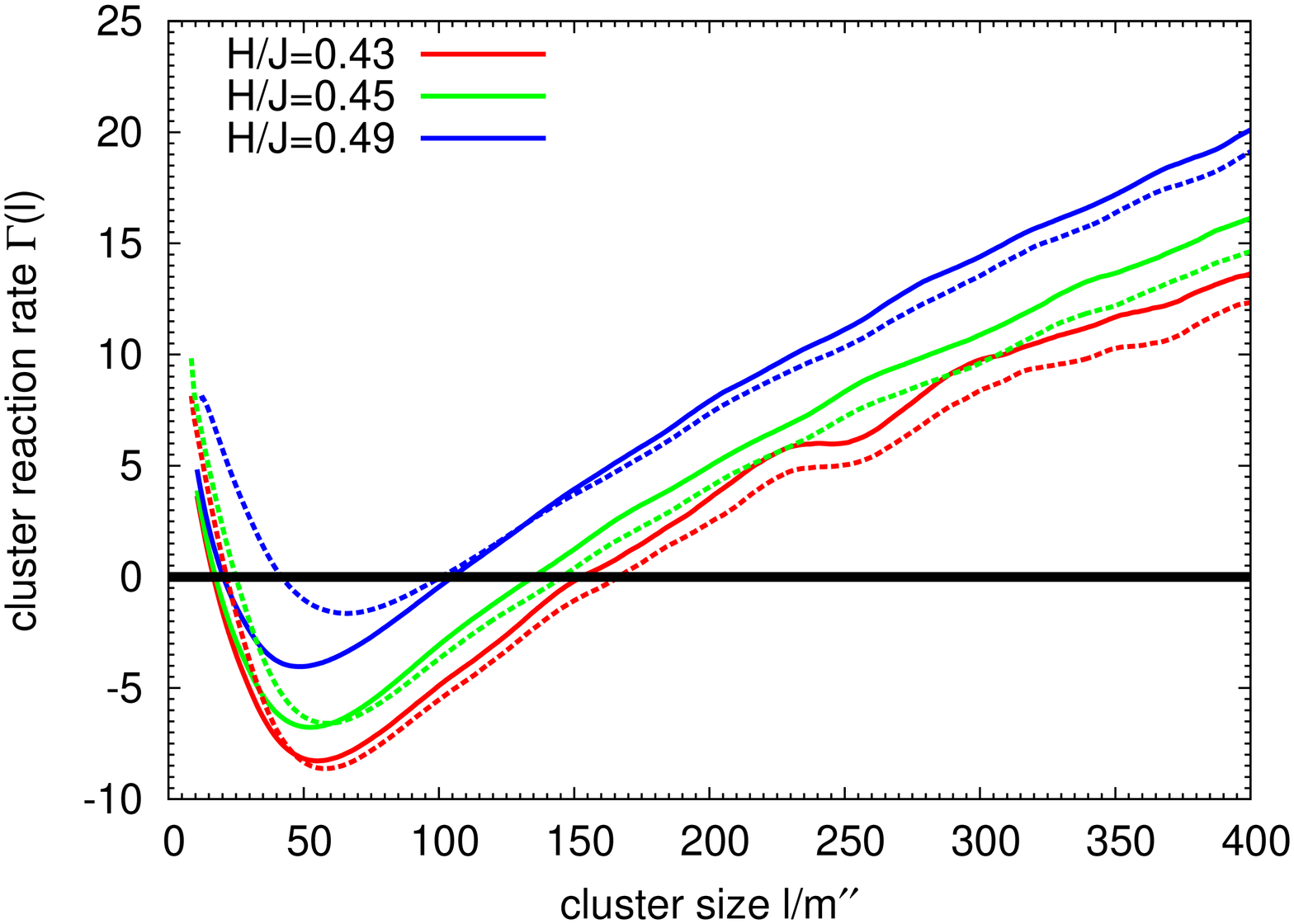}
	(b)
	\includegraphics[clip=true, trim=0cm 12mm 0mm 0cm, angle=-90,width=0.46 \textwidth]{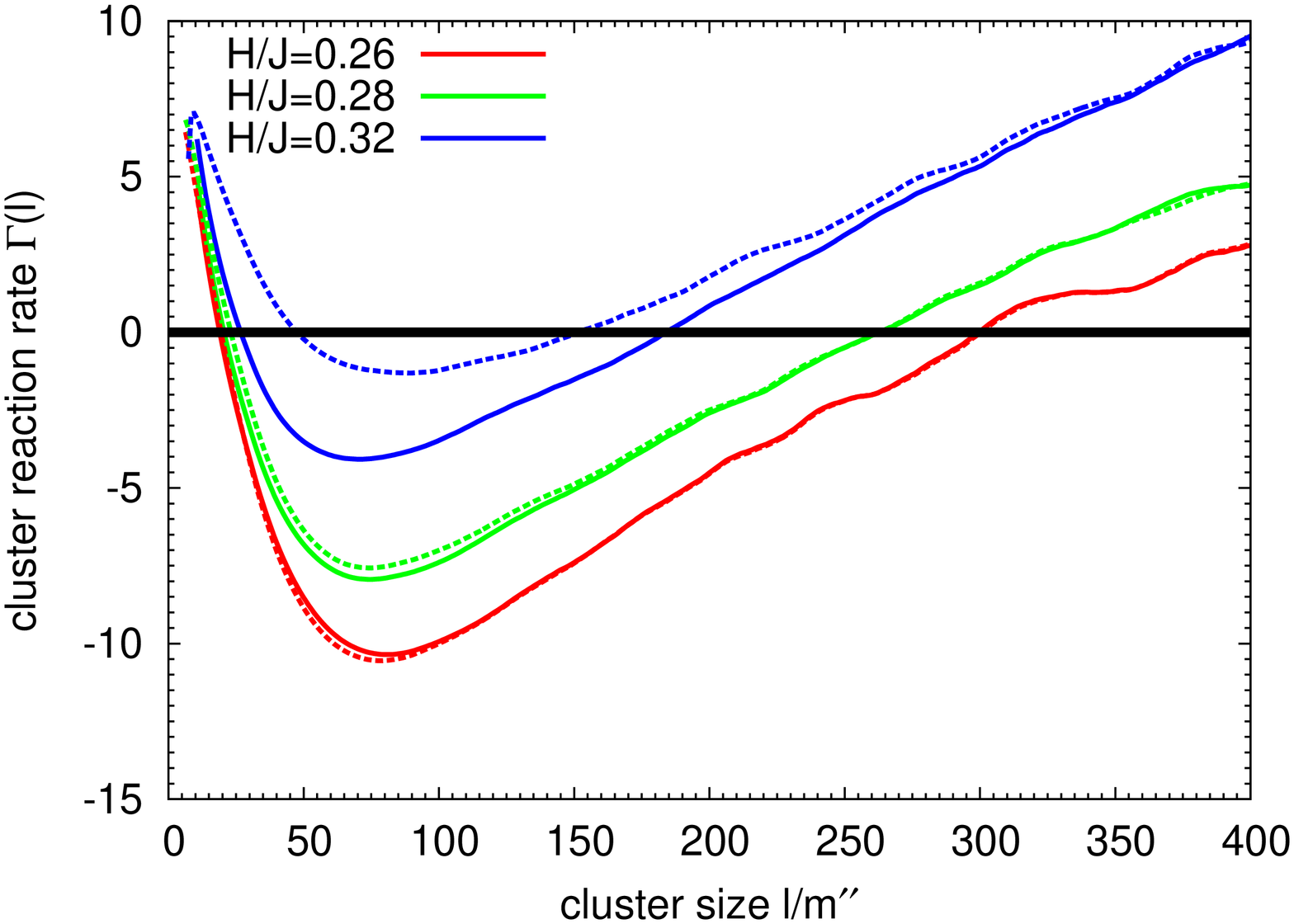}
	\caption{(Color online) Cluster reaction rate $\Gamma (\ell)$ versus cluster size $\ell$ for a bulk system at $k_BT/J=3.0$ without walls (a) and a system with walls at $k_BT/J=3.0$ and surface field $H_1/J=0.1$ (b). In each case several choices of the bulk field $H$ are shown, increasing from bottom to top, for clusters grow more likely with larger field strength. Full curves are based on Eq.~\eqref{eq17}, while broken curves are based on the approximation based on the use of the largest cluster only. In the heterogeneous case (b), both methods agree very well at lower fields because there is only one larger cluster in the system at a time.}
	\label{fig10}
	\end{center}
	\end{figure}

While Eq.~\eqref{eq17} is based on using all clusters $\ell_i(t) > \ell_\text{min}$ at each time $t$, one can simplify matters by restricting the analysis only to the trajectory of the biggest cluster in the system \cite{86}. When one does this, one ignores possible problems from the fact that from time to time the identity of the largest cluster changes. Fig.~\ref{fig10} shows now typical results for $\Gamma (\ell)$, using both this latter approximation and the method based on Eq.~\eqref{eq17}. Both methods yield similar trends, although they differ somewhat in detail (particularly in the case of homogeneous nucleation in the bulk).

	\begin{figure}[ht]
	\begin{center}
	(a)
	\includegraphics[clip=true, trim=0cm 12mm 8mm 0cm, angle=-90,width=0.7 \textwidth]{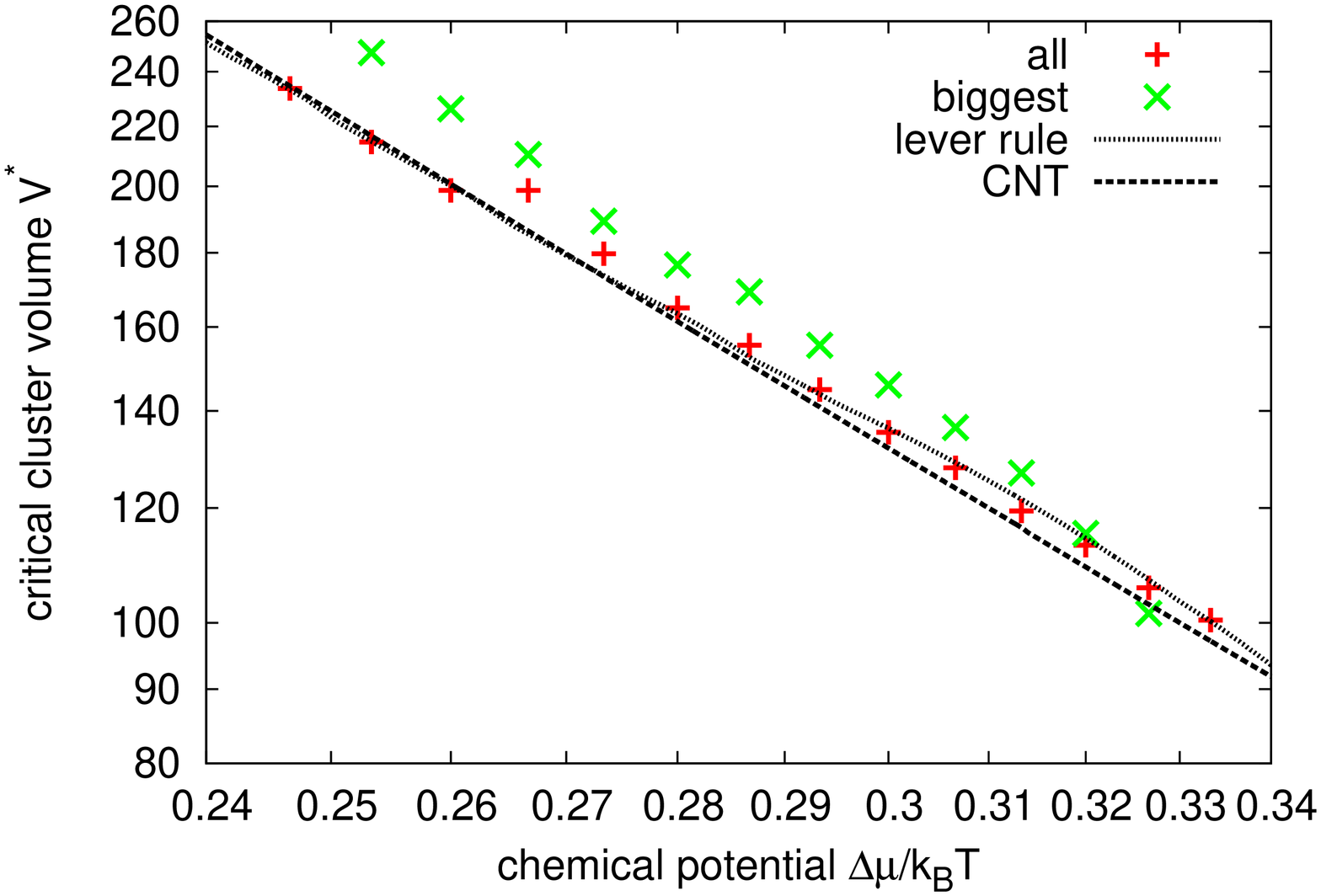}\\
	(b)
	\includegraphics[clip=true, trim=0cm 12mm 8mm 0cm, angle=-90,width=0.7 \textwidth]{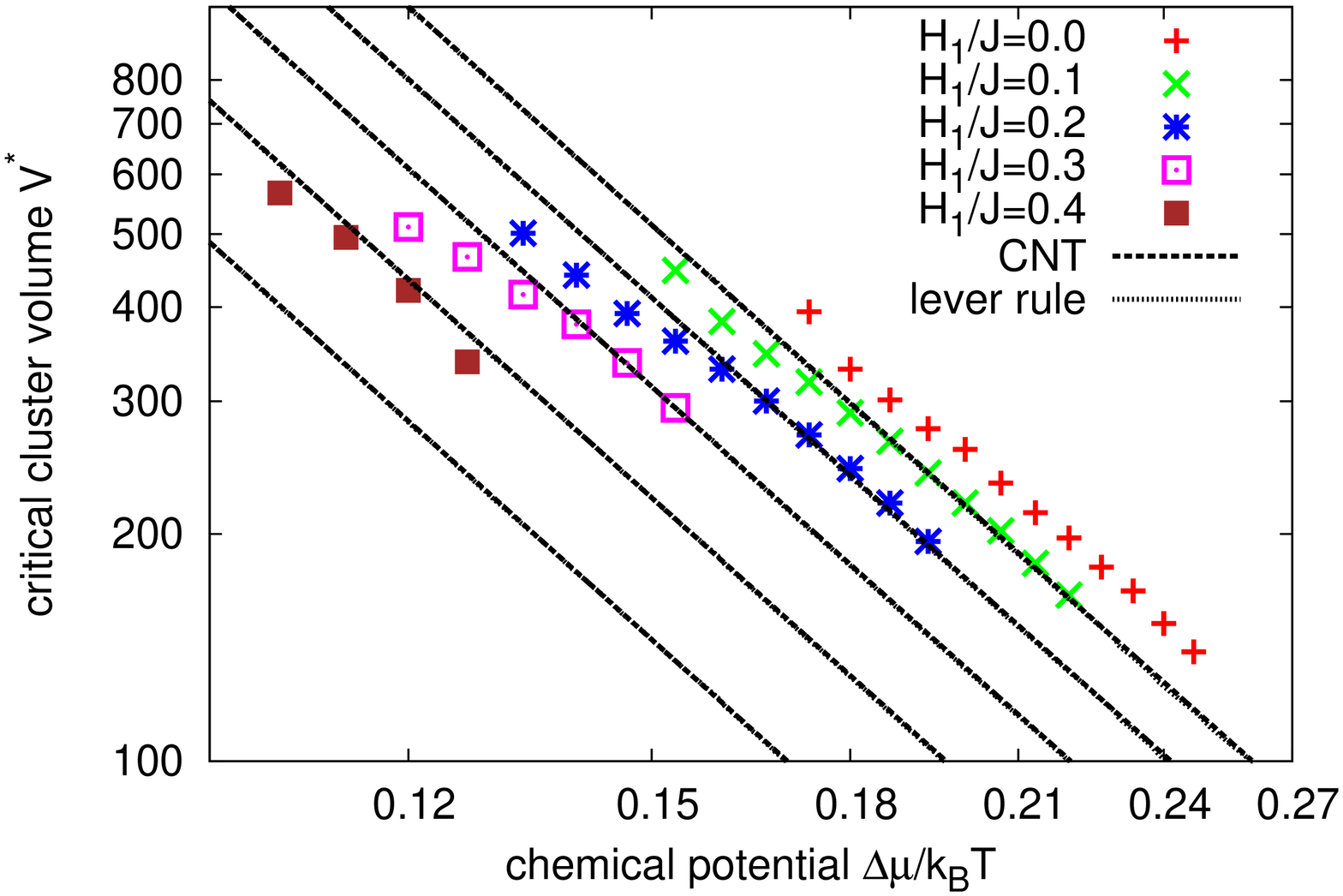}
	\caption{(Color online) Log-log plot of the critical cluster volume $V^*$ against the normalized chemical potential difference $\Delta \mu$ for $k_BT/J=3$ for homogeneous nucleation in the bulk (a) and heterogeneous nucleation at a wall with several choices of the surface field $H_1/J$, as indicated. The broken straight line is the prediction of the classical nucleation theory, Eq.~\eqref{eq3}, amended by the enhancement factor ($1.064$ at $k_BT/J=3$) taken from Fig.~\ref{fig: VCNTvsVLR}. The dotted lines are the corresponding results based on the leverrule. Results for the method based on Eq.~\eqref{eq17} are shown in both plots, the method based on the biggest cluster is only drawn in (a), as both methods yield the same results in the heterogeneous case. In case (a) the lever rule data were taken from a system at linear dimension $L=15$, and thus possibly affected by some finite size effects. Note also that in (b) the prediction of the classical nucleation theory and the leverrule results lie on top of each other.}
	\label{fig11}
	\end{center}
	\end{figure}

What one expects theoretically for $\Gamma (\ell)$ is a monotonous increase of $\Gamma (\ell)$ with $\ell$, where $\Gamma (\ell)$ is negative for clusters smaller than the critical cluster size $\ell ^*$ while $\Gamma (\ell)$ is positive for $\ell > \ell ^*$. The method yields a second positive part for $\ell$ slightly larger than $\ell_\text{min}$. This is an artifact of ignoring clusters smaller than $\ell_\text{min}$, which vanishes for $\ell_\text{min}\to 1$ \cite{86}.

If we identify the critical cluster size with the zero crossing of $\Gamma (\ell)$ at the right hand side, and convert to the cluster volume $V^*$ according to Eq.~\eqref{eq9}, we obtain the data shown in Fig.~\ref{fig11}. It is seen that for a given value of the field $H$ (or $\mu - \mu_\text{coex}$, respectively) classical nucleation theory underestimates the volume of the critical cluster: in other words, a given volume $V^*$ leads to a larger value of $H$. Since (according to classical nucleation theory and Eq.~\eqref{eq4a}) the barriers scale as $H^{-2}$, this means when one studies nucleation barriers as function of cluster volume $V^*$ or cluster radius $R^*$, one finds lower barriers in the simulation rather than predicted. Qualitatively, the data from the present analysis of cluster kinetics confirm the findings from the lever rule method of Winter et al. \cite{55}, as far as homogeneous nucleation is concerned (Fig.~\ref{fig11}a), although some questions on systematic errors in both methods have not been fully settled. Nevertheless, the qualitative agreement between these quite different approaches is satisfactory. For the case of heterogeneous nucleation, however, for a given supersaturation the critical cluster volume predicted from cluster kinetics (Fig.~\ref{fig10}) is distinctly larger than the corresponding results from the static methods. We have no explanation for this discrepancy.


\clearpage
\section{Concluding discussion}
 \label{sec4}

In the present work, we have studied aspects of nucleation theory by simulation of clusters and their dynamics, using the Ising (lattice gas) model on the simple cubic lattice. Both homogeneous nucleation and heterogeneous nucleation at planar walls (where a ``surface field'' acts) have been considered.

Although many aspects of this problem have been studied before in works of various groups extending over several decades, most of the previous work is inconclusive since it relied on the use of the ``geometric'' cluster definition. We have given evidence that this geometric cluster definition does not yield correct results for large clusters at the temperatures far above the roughening transition temperature where the clusters have spherical shape; at temperatures below the roughening transition temperature the geometric clusters and the ``physical clusters'' are basically indistinguishable, but due to the pronounced anisotropy effects a simple analysis of nucleation phenomena is not possible.

However, in the limit of large droplet volumes $V\to \infty$, where one can neglect any corrections to the decomposition of the droplet formation free energy into the bulk term plus a surface correction, one can compute the nucleation free energy barrier $\Delta F^*$ from measuring the excess chemical potential $\Delta \mu$ that is in equilibrium with a given $V$. Fig.~\ref{fig: VCNTvsVLR} shows the enhancement of $\Delta F^*$ with respect to the standard result for spherical droplets (using Eqs.~\eqref{eq3},~\eqref{eq4a}). We thus show that in the Ising (lattice gas) model this enhancement gradually rises from unity as the temperature is lowered from the critical temperature, reaches almost 10\% at $T/T_c=0.6$, and rises steeply below the roughening temperature towards the low temperature limit $6/\pi\approx 1.91$ (Fig.~\ref{fig: VCNTvsVLR}). This enhancement reflects the consequences of the anisotropy of the interfacial free energy, such as the gradual crossover of the droplet shape from a sphere to a cube (Fig.~\ref{fig6}).

On the other hand, we demonstrate that physical clusters do give consistent results, at least in the bulk when one is concerned with homogeneous nucleation. We show that in the limit where the droplets get macroscopically large, they converge against the simple lever rule predictions. However, we do find an (unexpected) surface excess in the particle number of such clusters also in this case. We also demonstrate the validity of the relation between chemical potential (of the supersaturated vapor) and the droplet radius that classical nucleation theory predicts for large droplets near the critical temperature. We also give evidence that the droplet-vapor interface is broadened due to capillary waves; we remind the reader that mean-field type theories and density functional theories \cite{11,14} cannot include such capillary wave effects (which also should give rise to a correction term on the droplet formation free energy, not yet included in Eq.~\eqref{eq2}).

We would also like to stress that many of our considerations can be carried over to a study of clusters in $d=2$ dimensions, where a construction as in Fig.~\ref{fig1} also holds. However, we expect two distinctions: (i) The roughening transition temperature $T_R$ is zero, so the crossover of droplet shape from the circle to the square occurs without any singularity even for arbitrarily large droplets. (ii) Percolation coincides with the critical point, but geometrical clusters still are too large, and to describe nucleation, physical clusters defined via Eq.~\eqref{eq4} should also be used. Of course, it would be very desirable to carry these considerations over to nucleation in off-lattice models of fluids. However, a precise analogue of Eq.~\eqref{eq4} is still not known, and hence other concepts to define physical clusters \cite{58} need to be used, if one wishes to study nucleation near the critical point.

In the second part we present a first study of the time evolution of the cluster population based on the ``physical cluster'' definition. However, due to the large computer resources needed for this study, only data at a single temperature ($k_BT/J=3.0$) are presented. In order to allow a comparison of this part of the study with our results on static properties of critical droplets, as studied in the first part of the paper, we use a criterion to estimate the critical droplet size from the balance between droplet growth and shrinking processes. In the case of homogeneous nucleation, the results obtained in this way are roughly compatible with the results obtained from the static lever rule method. Studying droplet volumes in the range from 100 to 200, clear deviations from the classical nucleation theory are seen, which can be attributed to a decrease of the nucleation barrier due to fluctuation effects. However, in the case of heterogeneous nucleation, a rather large discrepancy between the results of the statics and dynamics of droplets is found. This discrepancy is not understood yet, and must be left as a challenging problem for the future.

\textbf{Acknowledgements}: We thank D.~Winter for providing us with the data from Ref.~\cite{55} that were included in Fig.~\ref{fig11} for comparison. One of us (F.~S.) thanks the Deutsche Forschungsgemeinschaft for partial support under grant No VI 237/4-3 (SPP 1296).

 \end{document}